\documentclass[11pt]{article}

\usepackage{amsfonts}
\usepackage{amsmath}
\usepackage{graphicx}
\usepackage{stackrel}
\usepackage{rotating}
\usepackage{multirow}
\usepackage{helvet}
\usepackage{ccaption}
\usepackage{enumerate}
\usepackage{booktabs}
\usepackage{marvosym}
\usepackage{titletoc}
\usepackage{setspace}
\usepackage{bm}
\usepackage{url}
\usepackage{harvard}

\input{ee.sty}

\newtheorem{proposition}{Proposition}
\newtheorem{lemma}{Lemma}

\newenvironment{proof}{\paragraph{Proof}}{$\Box$}

\renewcommand{\baselinestretch}{1.3}
\title{\textbf{Semiparametric Conditional Quantile Models for Financial Returns and Realized Volatility}\footnote{We are indebted to Karim M. Abadir, Torben G. Andersen, Walter Distaso, Marcelo Fernandes, Simone Manganelli, Sergio Pastorello, anonymous referees and seminar participants at the 17th Annual Symposium of the Society for Nonlinear Dynamics and Econometrics (Atlanta, April 2009), North American Summer Meeting of the Econometric Society (Boston, June 2009), and Conference on Financial Econometrics and Realized Volatility/Vast Data (London, June 2009) for useful comments, suggestions and discussions. The first version of this paper was written while \v{Z}ike\v{s} was at Imperial College London; financial support from ESRC under grant RES-062-23-0311 is gratefully acknowledged. Barun\'{\i}k gratefully acknowledges financial support from the the Czech Science Foundation under project No. 13-32263S. The views expressed in this paper are those of the authors, and not necessarily those of the Bank of England.}}

\author{Filip \v{Z}ike\v{s}\thanks{Corresponding author: Bank of England, Financial Stability, Threadneedle street, London EC2R 8AH. Phone: +44 20 7601 5092. Email: filip.zikes@bankofengland.co.uk.} \and Jozef Barun\'{\i}k\thanks{Institute of Economic Studies, Charles University, Opletalova 21, 110 00, Prague,  CR and Institute of Information Theory and Automation, Academy of Sciences of the Czech Republic, Pod Vodarenskou Vezi 4, 182 00, Prague, Czech Republic.}}
\date{First version: 9 December 2010 \\ This version: 20 August 2013}

\begin{document}

\maketitle

\begin{abstract}
This paper investigates how the conditional quantiles of future returns and volatility of financial assets vary with various measures of ex-post variation in asset prices as well as option-implied volatility. We work in the flexible quantile regression framework and rely on recently developed model-free measures of integrated variance, upside and downside semivariance, and jump variation. Our results for the S\&P 500 and WTI Crude Oil futures contracts show that simple linear quantile regressions for returns and heterogenous quantile autoregressions for realized volatility perform very well in capturing the dynamics of the respective conditional distributions, both in absolute terms as well as relative to a couple of well-established benchmark models. The models can therefore serve as useful risk management tools for investors trading the futures contracts themselves or various derivative contracts written on realized volatility. \bigskip

\noindent \textbf{JEL:} C14, C21, G17, G32

\noindent \textbf{Keywords:} conditional quantiles, Value-at-Risk, quantile regression, realized measures
\end{abstract}

\newpage

\section{Introduction}
A fast growing recent literature in financial econometrics focuses on measuring, modeling and forecasting volatility using high-frequency data (\citename{abd09}, 2009). Yet, a number of important financial decisions require the specification and estimation of the entire distribution of future price changes and volatility, or at least of a few quantiles. Prime examples include portfolio selection when returns are non-Gaussian, risk measurement and management (Value-at-Risk), and market-timing strategies where the sign of future prices changes is to be predicted (\citename{chd06}, 2006). Forecasting the conditional distribution of future returns or its quantiles based on the use of intraday data and nonparametric measures of ex-post variation in asset prices has so far attracted much less attention than forecasting realized volatility. Notable exceptions include \citeasnoun{abdl03}, \citeasnoun{gl04} and \citeasnoun{cgk08}, who all combine time-series models for realized volatility with either parametric or nonparametric estimators of conditional distributions, and the recent contributions by \citeasnoun{bg09}, \citeasnoun{ss09} and \citeasnoun{mm10}, who base their predictive densities on parametric return-based volatility models.

This paper follows a different route and proposes to couple the flexible semiparametric quantile regression framework  with nonparametric measures of the various components of ex-post variation in asset prices to study the properties of conditional quantiles of daily asset returns and realized volatility, and forecast their future values. The use of quantile regression in financial econometrics is not new (\citename{kz96}, 1996, \citename{chu01}, 2001, \citename{em04}, 2004, \citename{ct08}, 2008), but to the best of our knowledge, it has not yet been applied in combination with realized volatility and related measures.

Our approach has a number of advantages. First, by relying on nonparametric measures of volatility we avoid making restrictive assumptions on the dynamics of the underlying conditional distributions. Second, by decomposing the overall ex-post variation in the prices process into the continuous (diffusion) and discontinuous parts (jumps), we are able to study the predictive power of these two components separately. Given the recent evidence on the predictive power of contemporaneous jumps for future volatility (\citename{abdi07}, 2007, \citename{cpr10}, 2010) and the finding of \citeasnoun{tt11} that prices and volatility tend to jump together seems to suggests that jumps may perhaps contain information about quantiles of future returns and volatility as well. Third, the semiparametric nature of quantile regression avoids confining attention to the relatively restrictive class of location-scale models (\citename{chu01}, 2001). Last but not least, our models are very simple to estimate yet capture, through the highly persistent realized volatility measures, the persistent dynamics of the conditional quantiles documented by \citeasnoun{em04} for equity returns.

In addition to the information contained in the historical high-frequency returns, we also investigate the predictive power of the (risk-neutral) expectations of future volatility embedded in options prices. The benefits of including implied volatility into the information set used for forecasting future volatility has been recently documented, among others, by \citeasnoun{gl07} and \citeasnoun{bcn10}. See also \citeasnoun{btz09}, who find the ability of the variance risk premium to forecast future medium-horizon stock returns.  \citeasnoun{chm05} show that volatility implied by foreign exchange options help to predict, albeit imperfectly, future distributions of spot exchange rates. \citeasnoun{ct08} obtain similar results for conditional quantiles of monthly equity index returns. Motivated by this empirical work, we include implied volatility as an additional covariate into the quantile regression models.

Besides modeling conditional quantiles of future returns, we propose simple models for the quantiles of future realized volatility. We follow \citeasnoun{abdi07} and \citeasnoun{bcn10} and consider a heterogeneous quantile autoregressive model (HQAR) with jumps and implied volatility. This model can be viewed as an extension of the heterogeneous autoregression, originally proposed by \citeasnoun{c09} for modeling the conditional mean of realized volatility, to conditional quantiles. A particular version of this model falls into the class of quantile autoregressions studied by \citeasnoun{kx06}.

Our empirical study of the S\&P 500 futures prices between January 1997 and June 2008 reveals some interesting features of the conditional distribution. First, we find that both realized as well as implied volatility possess significant predictive power for quantiles of future returns. Second, upon decomposing realized volatility into realized downside and upside semivariance (\citename{bnks10}, 2010), we find that it is almost exclusively downside semivariance that drives both left and right tail quantiles. Thus the past negative intraday returns contain more information about future quantiles than the positive ones and this effect is not subsumed by option-implied volatility. Finally, jumps play little role in forecasting quantiles of future returns.

Turning to models for realized volatility, we find that the heterogeneous quantile autoregressive model captures the time variation in conditional quantiles of daily realized volatility very well both in-sample as well as out-of-sample. The impact of contemporaneous realized and implied volatilities on future volatility quantiles is much higher in the far right tail of the distribution than in the left tail confirming the presence of a significant volatility-of-volatility effect documented by \citeasnoun{cmpp08} and \citeasnoun{bkpt09}. Similar to return quantiles, we document that recent realized downside semivariance possesses strong predictive power for future realized volatility quantiles, leaving almost no role for realized upside semivariance. Finally, the variation associated with jumps comes out insignificant in all models considered.

We complement our empirical analysis by applying the quantile regression models to the WTI Crude Oil futures contract. Oil futures prices exhibit substantially higher volatility and volatility of realized volatility than S\&P 500 which provides us with an opportunity to test our methodology on less well-behaved financial time series. We find that our quantile models for oil futures perform equally well in terms of their ability to deliver accurate quantile forecasts and find qualitatively similar results regarding the predictive power of the various components of the overall quadratic variation for forecasting quantiles of future returns and volatility.

To assess the relative performance of our linear quantile regressions, we use the Conditional Autoregressive Value at Risk (CAViaR) model of \citeasnoun{em04} and the ARFIMA-based lognormal-normal mixture of \citeasnoun{abdl03} as benchmarks. Overall, we find that neither of the models dominate in terms of performance uniformly across assets or quantiles. Putting realized measures into the CAViaR model does not drive out the other variables in the CAViaR equation completely and it improves its performance. The linear quantile regressions with realized measures, however, seem to perform no worse than the realized CAViaR. The ARFIMA-based lognormal-normal mixture delivers generally poorer unconditional coverage but it often exhibits lower loss at the same time. For multi-day realized volatility forecasts, we find that the linear quantile regression seems to perform better, especially in the right tail of the distribution.

The rest of the paper unfolds as follows. Section \ref{sec:theory} sets out the theoretical framework, while Section \ref{sec:lqr} discusses conditional quantile estimation by regression quantiles. In Section \ref{sec:me}, we study the implications of the measurement error induced by replacing the unobserved volatility components by their sample counterparts and provide sufficient conditions ensuring that the measurement error vanishes asymptotically. In Section \ref{sec:others} we briefly discuss a couple of alternative models for conditional quantiles that we use for comparison with our linear quantile regressions. Section \ref{sec:evaluation} describes the methods we employe to evaluate the performance of the conditional quantile models and Section \ref{sec:data} describes the data. Empirical application is carried out in Section \ref{sec:results} and finally Section \ref{sec:conclusion} concludes.

\section{Theoretical Framework}\label{sec:theory}
We assume that the logarithmic price process obeys an It\^{o} semimartingale
\begin{equation} \label{eq:X}
X_t = X_0 + \int_0^t \mu_s \mathrm{d}s + \int_0^t \sigma_s \mathrm{d}W_s + J_t,
\end{equation}
where $\mu$ is a predictable process, $\sigma$ is cadlag, $W$ is standard Brownian motion and $J$ is a finite-activity pure jump process,
\begin{equation*}
J_t = \sum_{j=1}^{L_t} \kappa_j,
\end{equation*}
where $L$ is a counting process and the $\kappa_j$'s are random variables governing the size of jumps. The process in equation (\ref{eq:X}) is very general and allows for rich dynamics. In particular, it accommodates stochastic volatility with possibly discontinuous sample paths (\citename{tt11}, 2011), the leverage effect characterized by negative correlation between volatility and price innovations (\citename{blt06}, 2006), time-varying jump intensity and jump sizes (\citename{cm02}, 2002), etc. We do not make any parametric assumptions about the respective processes when estimating the quantiles of the distribution of future returns but rely instead on reduced-form semi-parametric quantile regression models coupled with nonparametric measures of volatility and jumps variation.

Associated with the semimartingale in equation (\ref{eq:X}) is a quadratic variation process
\begin{eqnarray*}
QV_t &=& \int_0^t \sigma_s^2 \mathrm{d}s + \sum_{0 \leq s \leq t} (\Delta J_s)^2, \\
          &\equiv& IV_t + JV_t,
\end{eqnarray*}
where $IV_t$ is the integrated variance, that is, the part of $QV_t$ due to the continuous part of the log-price process and $JV_t$ is the jump variation due to the purely discontinuous part of $X_t$. As detailed by \citeasnoun{abdl03}, quadratic variation is a natural measure of variability in the logarithmic price and its individual components serve as important imputs into many asset pricing models.

When studying the conditional distribution of future returns, we separate the contribution of the two components of the quadratic variation process, i.e. the continuous part from the jump part. Recent evidence from the volatility forecasting literature (e.g. \citename{abdi07}, 2007, \citename{cpr10}, 2010) indicates that the two sources of variation in the asset price possess substantially different time series properties and affect future volatility in a different way. Anticipating that similar results obtain for the entire conditional distribution, we now describe an approach to disentangling the integrated variance from jump variation.

Suppose we obtain a sample of size $T(M+1)$, corresponding to $T$ days each having $M+1$ intraday observations. Define $\Delta_i X_t = X_{t-1 + (i+1)/M} - X_{t-1 + i/M}$ as the $i$-th intraday return on day $t$. A consistent estimator of the overall quadratic variation is provided by the well-known realized volatility, introduced into financial econometrics by \citeasnoun{ab98},
\begin{equation*}
RV_{t,M} = \sum_{i = 0}^{M-1} (\Delta_{i} X_t)^2,
\end{equation*}
with $RV_{t,M} \stackrel{p}{\rightarrow} IV_t + JV_t$ as $M \rightarrow \infty$. To estimate the integrated volatility, $IV_t$, in the presence of jumps, we employ the median realized volatility introduced by \citeasnoun{ads12}\footnote{Other methods proposed in the literature include \citeasnoun{b-ns04}, \citeasnoun{cpr10}, \citeasnoun{m09}.}:
\begin{equation*}
MedRV_{t,M} = \frac{\pi}{6-4\sqrt{3}+\pi}\left(\frac{M}{M-2}\right)\sum_{i=0}^{M-3}\mathrm{med}(|\Delta_{i} X_t|,|\Delta_{i+1} X_t|,|\Delta_{i+2} X_t|)^2
\end{equation*}
We can now define consistent estimators of $IV_t$ and $JV_t$, denoted by $IV_{t,M}$ and $JV_{t,M}$, respectively, as follows
\begin{eqnarray*}
IV_{t,M} &=& MedRV_{t,M}  \\
JV_{t,M} &=& RV_{t,M} - IV_{t,M}
\end{eqnarray*}

In addition to the $IV-JV$ decomposition of the overall quadratic variation, \citeasnoun{bnks10} recently propose to decompose the realized volatility and jump variation into the part associated with negative intraday returns and the part due to the positive intraday returns:
\begin{eqnarray*}
RS_{t,M}^{-} &=& \sum_{i = 0}^{M-1} (\Delta_{i} X_t)^2 1_{\{\Delta_{i} X_t < 0\}} \overset{p}{\longrightarrow} 0.5 IV_t + \sum_{t-1 \leq s \leq t} 1_{\{\Delta J_s < 0\}}(\Delta J_s)^2, \\
RS_{t,M}^{+} &=& \sum_{i = 0}^{M-1} (\Delta_{i} X_t)^2 1_{\{\Delta_{i} X_t > 0\}} \overset{p}{\longrightarrow} 0.5 IV_t + \sum_{t-1 \leq s \leq t} 1_{\{\Delta J_s > 0\}}(\Delta J_s)^2,
\end{eqnarray*}
In an empirical application, the authors find that the realized downside semivariance ($RS_{t,M}^{-}$) seems to be much more informative than the realized upside semivariance ($RS_{t,M}^{+}$) for the purposes of forecasting future volatility. Similar results have been recently obtained by \citeasnoun{ps09}.

\section{Linear Quantile Regression Models}\label{sec:lqr}
Having described the theoretical framework, we now propose simple linear semiparametric models for the quantiles of future returns and volatility.

\subsection{Models for returns}
We assume that the $\alpha$-quantile of the distribution of future returns, conditional on the information set $\Omega_t$, can be written as a linear function of the various components of the current and past quadratic variation and weakly exogenous variables,
\begin{equation} \label{eq:QR}
q_{\alpha}(r_{t+1}|\Omega_t) = \beta_0(\alpha) + \vbeta_v(\alpha)^{\prime} \vv_{t,M} + \vbeta_z(\alpha)^{\prime} \vz_t.
\end{equation}
where
\begin{eqnarray*}
r_{t+1} &=& X_{t+1} - X_t, \\
\vv_{t,M} &=& (QV_{t,M}^{1/2},QV_{t-1,M}^{1/2},...,IV_{t,M}^{1/2},IV_{t-1,M}^{1/2},...,JV_{t,M}^{1/2},JV_{t-1,M}^{1/2},...)',
\end{eqnarray*}
$\vz_t$ is a vector of weakly exogenous variables and $\beta_0(\alpha),\vbeta_v(\alpha),\vbeta_z(\alpha)$ are vectors of coefficients to be estimated.

The equation (\ref{eq:QR}) is a linear quantile regression proposed by \citeasnoun{kb78}. They show that the parameters can be estimated by minimizing the following objective function,
\begin{equation} \label{eq:QRmin}
QR_{T,M}(\vbeta(\alpha)) \equiv \frac{1}{T} \sum_{t=1}^{T} \rho_{\alpha} (r_{t+1} - \beta_0(\alpha) - \vbeta_v(\alpha)' \vv_{t,M} - \vbeta_z(\alpha)' \vz_t),
\end{equation}
where
\begin{equation*}
\rho_{\alpha}(x) = (\alpha - \mathbf{1}\{x<0\})x,
\end{equation*}
and $\vbeta(\alpha) = (\beta_0(\alpha),\vbeta_v(\alpha)',\vbeta_z(\alpha)')'$. Although the optimization problem does not admit a closed-form solution, relatively simple and computationally fast algorithms for finding the minimum are available, see \citeasnoun{pk97}. A potential problem that may arise in small samples is the so-called quantile crossing, i.e. the estimated quantiles are not guaranteed to be monotonic in $\alpha$. If this occurs, the recently developed approach due to \citeasnoun{chfg10} can be employed to establish monotonicity of the estimated quantiles. In our empirical applications reported later in the paper, quantile crossing never arises.

\subsection{Models for Realized Volatility}
Inspired by the success of the of the heterogenous autoregressive model (HAR) for realized volatility developed by \citeasnoun{c09} and extended by \citeasnoun{abdi07}, we write the conditional $\alpha$-quantile of the realized quadratic variation $RV_{t+1,M}$ as
\begin{equation}
q_{\alpha}(RV_{t+1,M}|\Omega_t) = \beta_0(\alpha) + \vbeta_{v1}(\alpha)'\vv_{t,M} + \vbeta_{v5}(\alpha)'\vv_{t,t-5,M} + \vbeta_{v22}(\alpha)' \vv_{t,t-22,M} + \vbeta_z^{\prime}(\alpha)' \vz_t \label{eq:hqar}
\end{equation}
where
\begin{equation*}
\vv_{t,t-k,M} = \frac{1}{k} \sum_{j=0}^{k-1} \vv_{t-j,M}
\end{equation*}
is the average $\vv_{t,M}$ over the past $k$ days, and as before $\vz_t$ a set of regressors. We call this model the heterogenous autoregressive quantile model (HARQ). Note that for a particular choice of regressors, namely $\vv_{t,M} = (RV_{t,M},RV_{t-1,M},...,RV_{t-k,M})'$ for some $k$, the model falls into the class of quantile autoregression (QAR) studied by \citeasnoun{kx06}, and the HARQ then simply becomes a restricted version of the QAR model. The general model in equation (\ref{eq:hqar}) is linear in parameters and hence estimation proceeds along the same lines as described in the previous subsection.

\section{Measurement Error Problem}\label{sec:me}
The quantile regression models proposed in the previous section are based on realized measures rather than the true, unobserved components of price variation. Asymptotic theory for the realized measures dictates that as the number of intraday observations grows without bound, the realized measures approach their unobserved counterparts and, equivalently, the measurement error associated with the realized measures approaches zero. Thus, under certain conditions it may be feasible to obtain, asymptotically, conditional quantiles for the true quadratic variation or any of its components. Whether or not this is desirable depends on the application at hand. If, for example, the objective is to estimate value-at-risk for variance swap positions, one need not worry about the measurement error problem, since here the goal is to estimate the quantiles of the realized volatility calculated at a fixed sampling frequency stipulated by the variance swap contract, i.e. $q_{\alpha}(RV_{t+1,M}|\Omega_t)$ for some fixed $M$. However, if the goal is to estimate the quantiles of future asset returns volatility, one needs to make sure that the impact of the measurement error vanishes so that one indeed obtains quantiles for the true quadratic variation, $q_{\alpha}(QV_{t+1}|\Omega_t)$, rather than realized volatility.

In this section, we provide sufficient conditions ensuring that the feasible objective function, $QR_{T,M}$, based on the realized measures converges in probability to the infeasible one, $QR_{T}$, based on the true unobserved components of quadratic variation, uniformly on the parameter space. If these conditions are satisfied we obtain, asymptotically, the desired quantiles of the quadratic variation, $q_{\alpha}(QV_{t+1}|\Omega_t)$, rather than the realized variance, $q_{\alpha}(RV_{t+1}|\Omega_t)$. The sufficient conditions depend on the properties of the measurement errors associated with the realized measures, which in turn depend on the behavior of the volatility and jump processes driving the logarithmic price, and on the relative rate of growth of $M$ and $T$.

To establish the asymptotic equivalence, we follow the double-asymptotic approach of \citeasnoun{cds11}, who study fully nonparametric estimators of conditional distributions of integrated variance using realized measures. In doing so, they establish some useful results regarding the rate of decay of moments of the measurement error associated with a number of realized measures. We extend these results to the case of realized volatility and median realized volatility in the presence of jumps and employ these to prove the asymptotic negligibility of the measurement error for the estimation of conditional quantiles.

We will need the following assumptions:

\begin{description}
\item[(A1)] The logarithmic price process follows (1) with $\mu_t \equiv 0$,
\item[(A2)] The volatility process $\{\sigma_t\}$ is a strong mixing with size $-2r/(r-2), r>2$ satisfying $\mathbb{E}[(\sigma_t^2)^{2(k+r)}] < \infty$, and the jump sizes satisfy $\mathbb{E}[\kappa^{2k}] < \infty$ for some $k \geq 2$.
\item[(A3)] The counting process $L_t$ is a Poisson process with strictly stationary intensity.
\end{description}

\noindent Assumption A1 specifies the data generating process. To simplify the proofs we assume that the drift is equal to zero. Assumptions A2 and A3 ensure that the moments of the measurement errors associated with $IV_{t,M}$ and $JV_{t,M}$ exist and decay sufficiently fast, as the following Lemma shows.
\begin{lemma}
Under assumptions A1-A2, $\mathbb{E}[|N_{t,M}^{(c)}|^k] = O(M^{-k/2})$. If, in addition, A3 holds then $\mathbb{E}[|N_{t,M}^{(d)}|^k] = O(M^{-k/2})$.
\end{lemma}
\begin{proof}
See Appendix.
\end{proof}

The first result of the Lemma is the same as in Lemma 1 of \citeasnoun{cds11}, who prove this for a number of different realized measures of integrated variance. Assuming, in addition, A3 allows us to establish similar result for the measure of jump variation based on the difference between realized variance and median realized variance. Given Lemma 1, we then have the following:

\begin{proposition}
Under assumptions (A1) - (A3), if $T^{\frac{2}{2k-1}}M^{-1/2} \rightarrow 0$ as $T,M \rightarrow \infty$ and $\Theta$ is a compact parameter space, then $\sup_{\vbeta \in \Theta}|QR_{T,M}(\vbeta) - QR_{T}(\vbeta)| \overset{p}{\rightarrow} 0$.
\end{proposition}
\begin{proof}
See Appendix.
\end{proof}

The proposition shows that the number of intraday observations ($M$) has to grow faster than a power of ($T$) for the contribution of the measurement error associated with the realized measures of integrated variance and jump variation to degenerate in the limit. How faster $M$ must grow depends on the the number of moments the volatility and jump processes possess. If all moments exist (i.e. $k = \infty$), we obtain the intuitive result that the contribution of the measurement error is driven entirely by discretization (finite $M$), i.e. it suffices to have $M \rightarrow \infty$ regardless of how fast this happens relative to $T \rightarrow \infty$. The reason we cannot establish this intuitive result for any $k \geq 2$ is due to the fact that the standard mean-value argument does not apply due to the non-differentiability of the objective function. To circumvent this problem, we have to ensure that $\sup_t |IV_{t,M}-IV_t|$ and $\sup_t |JV_{t,M}-JV_t|$ decay sufficiently fast, and this in turn depends on $k$ and the relative rate of growth of $M$ and $T$.

\section{Competing conditional quantile models}\label{sec:others}
To assess the relative performance of the linear quantile regression models proposed in this paper, we consider a couple of well-established benchmark models. Following the suggestions of the referees, we compare the return regressions with the CAViaR model proposed by Engle and Manganelli (2004), augmented by the various realized measures and option-implied volatility, and the lognormal-normal mixture of Andersen et al. (2003). We also use the latter model as a benchmark for the realized volatility quantile regressions.

\subsection{CAViaR}
Engle and Manganelli (2004) propose a dynamic non-linear quantile regression model, the so-called CAViaR, for daily asset return quantiles, $q_t(\vtheta)$, where $\vtheta$ is a vector of parameters to be estimated. They consider four different specifications of $q_t(\vtheta)$, two of which we employ here:

Symmetric absolute value:
\begin{equation}
q_{t+1}(\vtheta) = \beta_1 + \beta_2 q_{t}(\vtheta) + \beta_3 |r_{t}| + \vgamma'\vx_{t},
\end{equation}

Asymmetric slope:
\begin{equation}
q_{t+1}(\vtheta) = \beta_1 + \beta_2 q_{t}(\vtheta) + \beta_3 (r_{t})^+ + \beta_4 (r_{t})^- + \vgamma'\vx_{t-1},
\end{equation}
where $(r_{t})^+ = r_t \mathbf{1}\{r_t \geq 0\}$ and $(r_{t})^- = r_t \mathbf{1}\{r_t < 0\}$. Two things are novel in our application of the CAViaR model. First, we include the various realized measures and implied volatility used in the linear regressions into the CAViaR equations, calling the augmented model realized CAViaR. The idea is that these variables are much better proxies for the past return volatility than the absolute return and should therefore improve the predictive performance of the baseline CAViaR model with $\vgamma \equiv \vzeros$. Since the realized measures and the option-implied volatility are significantly more persistent than the absolute return, including them into the model might also reduce or completely drive out the affect of the lagged quantile, $q_{t}(\vtheta)$.

Second, we use the realized CAViaR model to forecast not only daily returns, but also to 5-day and 10-day returns. We employ the direct forecasting approach whereby we fit the model to the 5-day and 10-day returns directly, rather than using the model for 1-day returns to generate 5-day and 10-day quantile forecasts. That way, the multi-day forecasts can be obtained directly from the realized CAViaR equations and we do not have to write down and estimate separate equations for the various lagged variables entering the CAViaR recursion. To the best of our knowledge, this is the first application of the CAViaR model to multi-day quantile forecasting.

Similarly to the linear quantile regressions, the realized CAVIaR can be estimated by minimizing the check function given by
\begin{equation}
Q_T(\vtheta) = \frac{1}{T} \sum_{t=1}^T (\alpha - \mathbf{1}\{r_t < q_t(\vtheta)\})(r_t - q_t(\vtheta)).
\end{equation}
However, due to the nonlinear nature of the model, no simple algorithm for this optimization problem exists and we resort to the fairly elaborate procedure proposed by Engle and Manganelli (2004). Computing standard errors for the CAViaR parameter estimates requires a choice of bandwidth (see Engle and Manganelli, 2004) and there is currently no procedure available for the optimal choice. We proceed by calculating standard errors for a range of bandwidth values, select a region where the standard errors are relatively stable and report standard errors corresponding to a bandwidth from this region.\footnote{We are grateful to Simone Manganelli for suggesting this approach.}

\subsection{Long-memory lognormal-normal mixture}
Our second benchmark for the return models and a benchmark for the realized volatility models is the lognormal-normal mixture model proposed by Andersen et al. (2003):
\begin{eqnarray}
r_t &=& RV_{t,M}^{-1/2} \epsilon_t, \\
(1-\phi L)(1-L)^d \log RV_{t,M} &=& (1-\psi L)u_t
\end{eqnarray}
where $\epsilon_t$ is \emph{iid} standard normal and $u_t$ is \emph{iid} $\mathrm{N}(0,\sigma_u^2)$ independent of $\epsilon_t$. In this model, the logarithmic realized volatility follows a Gaussian ARFIMA(1,d,0) process so that realized volatility is unconditionally lognormally distributed, while returns are conditionally Gaussian and unconditionally mixed-Gaussian.

We fit the model to daily returns and realized volatilities using maximum likelihood. One-day ahead quantile forecasts for returns and realized volatility can be obtained analytically, but multi-day forecasts have to be simulated since the distribution function of a sum of lognormal random variables is not available in closed form.

\section{Evaluation of quantile forecasts}\label{sec:evaluation}
We evaluate the \emph{absolute} performance of the various conditional quantile models using the CAViaR test of \citeasnoun{bcp11}, which is a version of the DQ test of \citeasnoun{em04}. In particular, we define a ``hit" variable
\begin{equation*}
Hit_{t+1} = \mathbf{1}\{r_{t+1} \leq q_{\alpha}(r_{t+1} | \Omega_t)\},
\end{equation*}
which is a binary variable taking on the value of one if the conditional quantile is violated and zero otherwise. If the conditional quantiles are correctly dynamically specified, the sequence of hits should be \emph{iid} Bernoulli distributed with parameter $\alpha$. To test this hypothesis, \citeasnoun{bcp11} propose to estimate the following logistic regression
\begin{equation}\label{eq:berkowitz}
Hit_{t} = c + \sum_{k=1}^n \beta_{1k} Hit_{t-k} + \sum_{k=1}^n \beta_{2k} q_{\alpha}(r_{t-k+1} | \Omega_{t-k})\} + u_t.
\end{equation}
and use the likelihood ratio test for the null hypothesis that the $\beta$ coefficients are zero and $\mathbb{P}(Hit_t = 1) = e^c/(1+e^c)=\alpha$. We use Monte Carlo simulation to obtain exact finite-sample critical values for the likelihood ratio test as suggested by \citeasnoun{bcp11}.

This approach to evaluating absolute performance of quantile forecasts is only suitable for one-step-ahead forecasts. To see this, define the $h$-period hits as
\begin{equation}
Hit_{t|t+h} = \mathbf{1}\{r_{t+1}+r_{t+2}+\cdots+r_{t+h} \leq q_{\alpha}(r_{t+1}+r_{t+2}+\cdots+r_{t+h}  | \Omega_t)\},
\end{equation}
where $q_{\alpha}(r_{t+1}+r_{t+2}+\cdots+r_{t+h}  | \Omega_t)$ is the quantile forecast for the cumulative $h$-period return given the information available at time $t$. Clearly, even if the quantiles are dynamically correctly specified, the sequence of hits $\{Hit_{t|t+h}\}$ is $h$-dependent, which violates the assumptions underlying the likelihood ratio test in the logit model in equation (\ref{eq:berkowitz}). A solution to this problem could be to test the null hypothesis in an OLS regression of $Hit_{t|t+h}$ on a constant and $Hit_{t|t-jh}$, $j=1,...,n$, using a Wald test statistic with the Newey-West variance. The latter would account for both heteroskedasticity and serial correlation in the regression. We have experimented with this approach in a Monte Carlo simulation (available on request) and find that while it works well in very large samples as dictated by asymptotic theory, the finite-sample performance of the test is poor: the test is heavily oversized even with 1,000 observations. To the best of our knowledge, there is currently no alternative, reliable test for correct dynamic specification of multi-step conditional quantiles.

To assess the \emph{relative} performance of the various quantile models, we follow \citeasnoun{cgk08} and focus on pairwise comparison based on the tick-loss function suggested by \citeasnoun{gk05}:
\begin{equation}\label{eq:tickloss}
L_{\alpha}(e_{t+1}) = (\alpha - 1\{e_{t+1}<0\})e_{t+1},
\end{equation}
where $e_{t+1} = r_{t+1} - q_{\alpha}(r_{t+1} | \Omega_t)$. To test for equal predictive ability we use the \citeasnoun{dm95} test with the Newey-West variance in the case of multi-step-ahead quantile forecasts.

\section{Data Description and Preliminaries}\label{sec:data}
We apply the conditional quantile models to returns and realized volatility of two assets: S\&P 500 and WTI Crude Oil futures.

We use high-frequency data on the S\&P 500 futures contract obtained from Tick Data for a period running from January, 1996 till June, 2008. We focus on transactions prices pertaining to the most liquid (front) contract traded on the Chicago Mercantile Exchange (CME) during the main trading hours of 9:30 - 16:00 EST. From the raw irregularly spaced prices we extract 5-minute logarithmic returns using the last-tick method. The choice of sampling frequency is guided by the volatility signature plot (\citename{abdl00}, 2000), and previous literature employing the same data (\citename{abdi07}, 2007, \citename{cpr10}, 2010, among others).

In addition to historical volatility measures, we also explore the role of option implied volatility. In particular, we employ the VIX index calculated by the Chicago Board of Exchange (CBOE), which measures market expectations of one-month-ahead volatility of the S\&P 500 index implied by a portfolio of put and call options. The index is model-free in the sense that it does not rely on any particular parametric option pricing model to extract the implied volatility. \citeasnoun{fms09} provide a detailed description of the construction of the index as well as its time-series properties. Although the maturity of the options used to construct the index (30 calendar days) does not match our forecasting horizons, the VIX index can still be used, and very successfully as we will see later, as a proxy for future volatility.

The intraday WTI Crude Oil futures prices are obtained from Tick Data and cover the period from September, 2001 till August 2008. Similar to the equity futures, we focus on the front contract traded on the New York Mercantile Exchange (NYMEX) during the main trading hours between 9:00 - 15:00 EST. We employ 5-minute logarithmic returns to avoid issues with market microstructure noise.

The CBOE has recently introduced a crude oil volatility index (OVX), applying the same methodology as in the case of VIX to calculate 30-day volatility implied by oil futures options. The history of OVX only goes back to May 2007 and so is too short for our purposes. We therefore construct our own model-free implied volatility index using settlement prices for American-style futures options on oil traded on the CME, following the methodology of \citeasnoun{cw09} and \citeasnoun{ts09}. The details are described in Appendix B.

\subsection{Returns, realized measures and implied volatility: S\&P 500 futures}
\begin{figure}[!h]
\centering
\begin{tabular}{cc}
  \includegraphics[width=75mm,height=65mm]{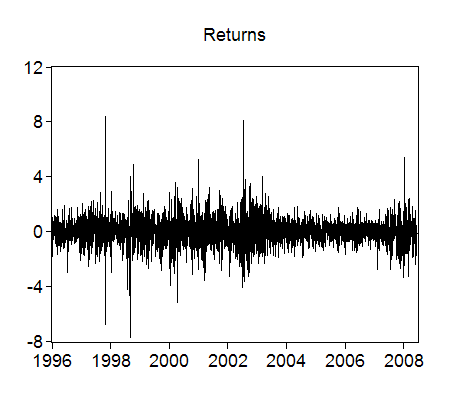} & \includegraphics[width=75mm,height=65mm]{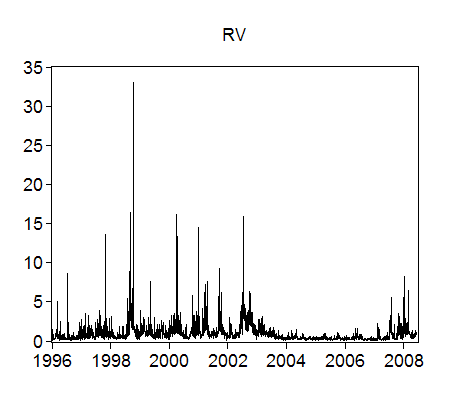} \\
  \includegraphics[width=75mm,height=65mm]{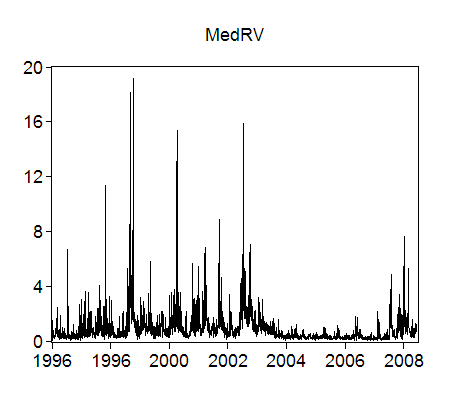} & \includegraphics[width=75mm,height=65mm]{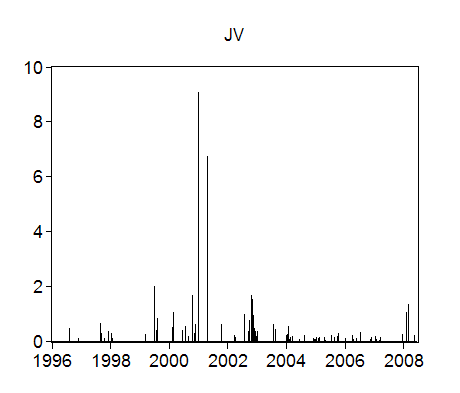}
\end{tabular}
\caption{Time series of daily returns, realized variance, median realized variance and jump variation for the S\&P 500 futures contract. All realized measures are calculated from 5-minute prices obtained from irregularly-spaced transactions data using the last-tick method. The sample period is from January 3, 1997 till June 30, 2008.\label{fig:dataSP}}
\end{figure}

We construct the following measures of the various components of quadratic variation: realized variance, realized upside semivariance, realized downside semivariance and the median realized volatility. As mentioned before, the median realized volatility offers a number of advantages over the alternative measures of integrated variance in the presence of infrequent jumps. It is less sensitive to the presence of occasional zero intraday returns and enjoys smaller finite-sample bias induced by jumps, while being computationally simple to implement. Table \ref{tab:statssp} reports the summary statistics for the daily open-to-close logarithmic returns and the various measures of variation in the S\&P 500 futures prices. The daily returns, plotted in Figure \ref{fig:dataSP}, exhibit the usual stylized properties of financial returns: small, insignificant mean, excess kurtosis and volatility clustering.

Turning to the realized variance and the upside and downside semivariances, we observe that they are all highly positively skewed. A logarithmic transformation does not eliminate the skewness entirely leading to the rejection of normality of logarithmic RV and hence log-normality of the realized variance and semivariances. The realized upside variance seems to be slightly more volatile than the realized downside variance and its distribution is also much more positively skewed and heavy-tailed. The Ljung-Box test for no autocorrelation up to lag 20 confirms the well-known long-memory features of realized volatility.

To estimate the contribution of jumps, we first test on a day-by-day basis for the presence of jumps in the price process using a test based on the median realized volatility\footnote{Although \citeasnoun{ads12} do not derive a test for jumps based on $MedRV$, this can be easily done by exploiting their joint Central Limit Theorem for $RV$ and $MedRV$ and following the steps of \citeasnoun{b-ns06}. Simulation evidence reported by \citeasnoun{tz09} indicates that a test based on the ration of $MedRV$ and $RV$ enjoys good finite sample properties and some robustness to the presence of occasional zero intraday returns.}. We set the significance level to 0.1\% as is usual in the literature. On days when jumps are detected by the test, we set $IV_{t,M} = MedRV_{t,M}$ and $JV_{t,M} = RV_{t,M} - MedRV_{t,M}$, while on days when no jumps are found, we set $IV_{t,M} = RV_{t,M}$ and $JV_{t,M} = 0$, thereby ensuring that the continuous and discontinuous components always sum up to the overall quadratic variation. This shrinkage approach follows, among others, \citeasnoun{abdi07} and \citeasnoun{cpr10}.

Similar to previous empirical results (\citename{ht05}, 2005) we find that jumps are relatively infrequent. The test identifies 66 days with significant jumps corresponding to about 2.1\% of days in our sample. The jumps contribute only about 1.3\% to the overall quadratic variation. It is clear from the plot of the time series of jump variation (Figure \ref{fig:dataSP}) that the properties of jumps have changed roughly in the middle of the sample period. While over the first 5-6 years of the sample the jumps were rare and large, it seems that they have become smaller and more frequent in the second half of our sample period. Note that this period is also associated with relatively small integrated variance as measured by the median realized variance.

Finally, we look at the properties of the VIX index. The Ljung-Box $Q$ statistic indicates high degree of persistence, much higher than for the realized measures of ex-post variance. The VIX implied volatility, however, pertains to a 30-calendar-day period and hence the daily observations involve a great degree of overlap. It is thus not surprising to find such high and slowly decaying autocorrelation. Note also that the mean implied volatility is larger than the mean realized volatility, confirming the existence of a negative variance risk premium, see e.g. \citeasnoun{btz09} and the references therein for more evidence.

\subsection{Returns and realized measures and implied volatility: Crude oil futures}
\begin{figure}[!h]
\centering
\begin{tabular}{cc}
  \includegraphics[width=75mm,height=65mm]{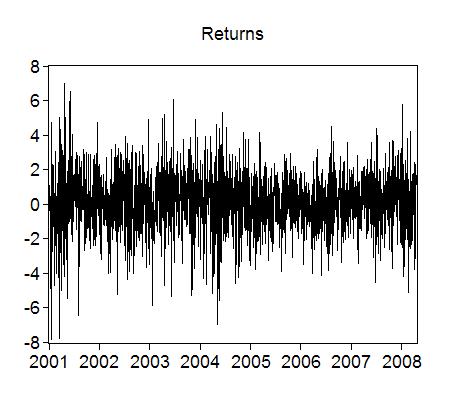} & \includegraphics[width=75mm,height=65mm]{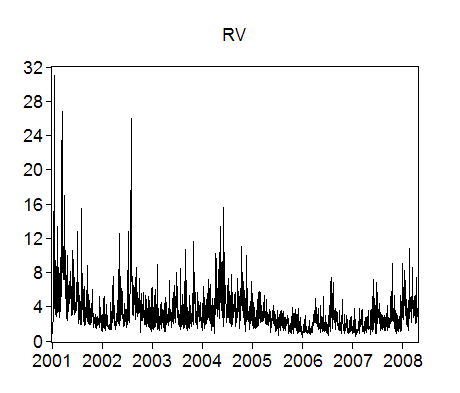} \\
  \includegraphics[width=75mm,height=65mm]{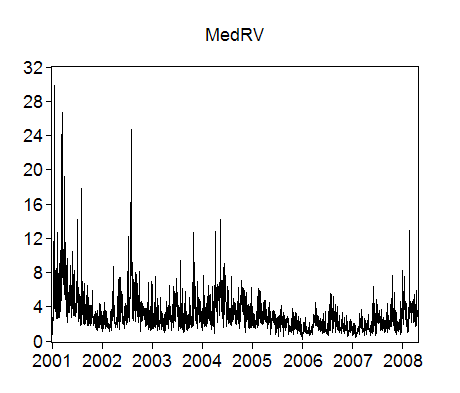} & \includegraphics[width=75mm,height=65mm]{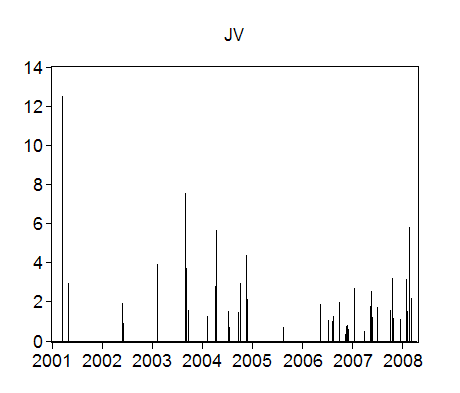}
\end{tabular}
\caption{Time series of daily returns, realized variance, median realized variance and jump variation for the WTI Crude Oil futures contract. All realized measures are calculated from 5-minute prices obtained from irregularly-spaced transactions data using the last-tick method. The sample period is from September 4, 2001 till August 30, 2008.\label{fig:dataOil}}
\end{figure}

We now repeat the same exercise with the WTI Crude Oil futures prices. The summary statistics for daily returns and the various realized measure are reported in Table \ref{tab:statsoil} and their time-series are plotted in Figure \ref{fig:dataOil}. We observe that the daily oil futures returns are highly volatile, with the average daily realized variance at about 4\% exceeding the average $RV$ of S\&P 500 by more than four times. The volatility of realized volatility is also substantially larger, while the Ljung-Box test statistics indicates smaller degree of serial correlation. That the oil futures realized volatility is highly volatile and relatively less persistent is also apparent from the time-series plot depicted in Figure \ref{fig:dataOil}. All realized measures exhibit positively skewed and heavy-tailed unconditional distributions.

Similar to \citeasnoun{ts09} we find that the model-free implied volatility is, on average, higher than realized volatility, confirming the existence of priced variance risk in the oil market. The magnitude of the variance risk premium is smaller than in the equity market. Applying the test for jumps on a day-by-day basis we identify 38 days when the oil futures price jumped by a significant amount, corresponding to 2\% of days in the sample. The estimated contribution of jumps to the total variation is about 1.5\%. Figure \ref{fig:dataOil} shows that the jumps are relatively large and rare.

\section{Empirical Results}\label{sec:results}
\subsection{Return quantiles}
\subsubsection{Estimation and in-sample fit}
We begin by modeling and forecasting quantiles of daily returns, focusing on the 5\%, 10\%, 90\% and 95\% quantiles and the median since these are most interesting from an economic point of view. Throughout, we employ realized volatilities rather than variances, i.e. we take the square root of the realized measures discussed above. Estimation of linear quantile regressions is carried out using the interior-point method of \citeasnoun{pk97} and standard errors are obtained by moving-block bootstrap (\citename{f97}, 1997). For CAViaR models we use the estimation approach of Engle and Manganelli (2004). The ARFIMA models for logarithmic realized variances is estimated by maximum likelihood.\footnote{Linear quantile regressions are estimated using the RQ package for Ox Version 1.0 developed by \citeasnoun{pk97}. ARFIMA models are estimated by the ARFIMA package 1.04 for Ox by \citeasnoun{do06}. CAViaR models are estimated using the MATLAB and C++ routines by Simone Manganelli, adapted to accommodate weakly exogenous variables.}

A large number of different specifications of the quantile regression models can be considered. To save space, we only report models that provide interesting insights into the dynamics of conditional quantiles while at the same time deliver accurate out-of-sample quantile forecasts. The estimation results are reported in the upper panels of Tables \ref{tab:qrsp} and \ref{tab:qroil} for S\&P 500 and WTI Crude Oil futures returns, respectively. We first discuss results for the upper and lower tail quantiles and the median separately as the latter are very different from the former. \medskip

\noindent \emph{Lower and upper tail conditional quantiles}

\noindent For both assets, we find that the lagged realized volatility is highly statistically significant in the linear quantile regressions (LQR) across the different quantiles. The estimated parameter have the expected sign: the left-tail (right-tail) quantiles vary negatively (positively) with realized volatility. Turning to the symmetric absolute value (SAV) CAViaR model, we find qualitatively similar parameters estimates as Engle and Manganelli (2004) in that the lagged conditional quantile parameter is close to one and highly statistically significant, while the lagged absolute return coefficient is relatively small but also significant. Including the lagged realized volatility into the CAViaR equation (Realized CAViaR) reduces the coefficient associated with the lagged conditional quantile, but only slightly and without affecting its statistical significance. In case of the S\&P 500 futures, the lagged realized volatility drives out the lagged absolute return in the lower-tail quantiles, but both variables remain statistically significant in the upper-tail quantiles. In case of the WTI Crude Oil, neither lagged $RV$ nor absolute return turn out to be statistically significant at the 5\% level, owing perhaps to collinearity, though the lagged realized volatility tends to command higher parameter estimates and t-statistics (in absolute value) than the lagged absolute return.

Next, we decompose the realized variance into the continuous and jump parts and estimate quantile regressions in which the measures of integrated variance and jump variations enter separately. We also add the option-implied volatility into the conditional quantile equations. The estimation results are reported in the middle panels of Tables \ref{tab:qrsp} and \ref{tab:qroil}. We find that jumps play essentially no role in the linear quantile regressions (LQR) as $JV$ turns out to be statistically insignificant across the board. Lagged integrated volatility comes out highly significant in the S\&P 500 regressions but insignificant in the WTI Crude oil regression. This is perhaps due to the effect of the option-implied volatility that clearly plays a major role in the conditional quantiles of both asset returns; the associated parameter estimates are relatively large in magnitude and highly significant.

Adding the $IV$, $JV$ and implied volatility into the SAV CAViaR model (Realized CAViaR) produces different results across the two assets. In case of S\&P 500, we find that the coefficient of the lagged conditional quantile is now substantially reduced, perhaps due to the strong predictive power of implied volatility, and becomes statistically insignificant in the lower tail. The lagged integrated volatility remains statistically significant in both tails, while the lagged VIX only in the lower tail. Interestingly, the lagged absolute return is not driven out in the upper tail, although the associated coefficient estimates are counter-intuitively negative. In case of WTI Crude Oil, we find that neither the lagged absolute return nor the lagged integrated variance come out significant, while option-implied volatility only appears to matter in the 95\% quantile. Rather surprisingly, the jump variation becomes significant in the Realized CAViaR; the estimated coefficients have the right sign and are relatively large in magnitude.

\begin{figure}[!h]
\centering
\captionnamefont{\bf}
\begin{tabular}{cc}
  \includegraphics[width=75mm,height=68mm]{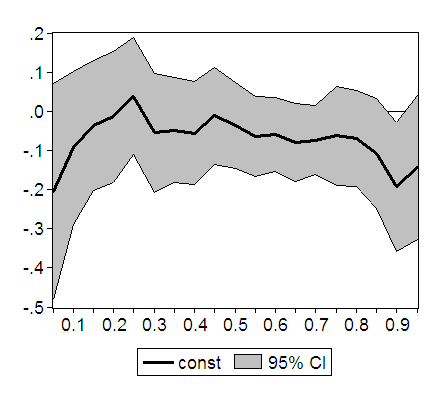} & \includegraphics[width=75mm,height=68mm]{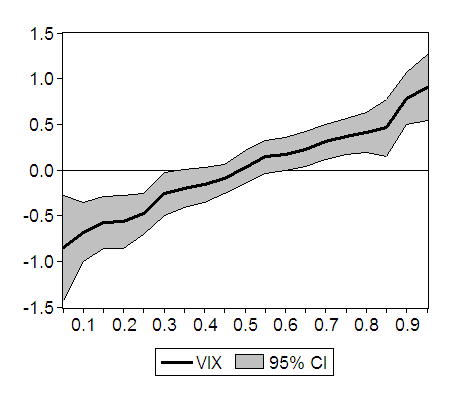} \\
  \includegraphics[width=75mm,height=68mm]{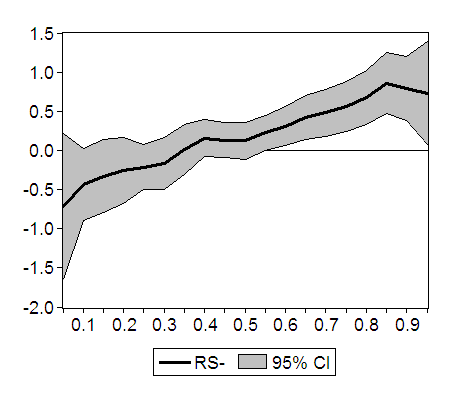} & \includegraphics[width=75mm,height=68mm]{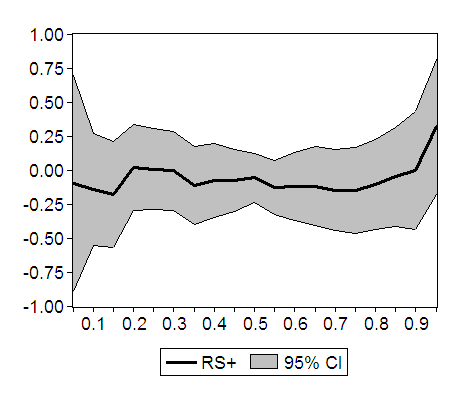}
\end{tabular}
\caption{Estimated quantile regression process for model LQR3 in Table \ref{tab:qrsp} for S\&P 500 futures returns. For each $\alpha$-quantile ranging from 0.05 to 0.95, we plot the estimated parameters in the quantile regression $(\widehat{\vbeta}(\alpha))$ together with pointwise 95\% bootstrapped confidence intervals.\label{fig:qrpSPret}}
\end{figure}

Finally, we decompose the realized variance into upside and downside semivariances and allow these to enter the quantile regressions separately. We also include the option-implied volatility. The lower panels of Tables \ref{tab:qrsp} and \ref{tab:qroil} report the estimation results and Figure \ref{fig:qrpSPret} illustrates the results graphically for a wider range of quantiles. The realized downside volatility clearly dominates across all estimated quantiles and leaves virtually no role for the upside volatility in the linear quantile regression. The information content of the downside volatility is not subsumed by option-implied volatility, which itself turns out to be highly statistically significant.

These results are consistent with the estimates of the asymmetric slope (AS) CAViaR, where only the coefficient associated with the lagged negative return are generally statistically significant, as in Engle and Manganelli (2004). Adding the realized semivariances and option-implied volatility into the AS CAViaR equations produces mixed results: the parameter estimates tend to be insignificant and do not always have the expected sign. Our conjecture is that this may be due to collinearity.

Having discussed the estimation results we now turn to evaluating the in-sample fit of the alternative daily conditional quantile models using the methodology of \citeasnoun{bcp11} as described in Section 6. The results are summarized in left-hand side panels of Tables \ref{tab:insampleretSP} and \ref{tab:insampleretOil}. For each model and quantile, we report the in-sample unconditional coverage ($\hat{\alpha}$), the likelihood ratio test statistic ($DQ$) for the null hypothesis that all the beta's in the logistic regression (\ref{eq:berkowitz}) are equal to zero and the associate Monte Carlo-based p-value ($p$-val). We run the logistic regressions with 5 lags.

Starting with S\&P500 futures we find that all models perform very well in the lower tail, having the unconditional coverage very close to the nominal levels and comfortably passing the \citeasnoun{bcp11} test. Some dynamic misspecification is indicated by the test in the upper-tail quantiles for ARFIMA and the symmetric absolute value CAViaR models with and without realized measures, especially for the 90\% quantile. The DQ test also rejects the correct specification of this quantile for the linear quantile regression with lagged realized volatility as the only regressor (LQR1). The asymmetric CAViaR specifications as well as the linear quantile regressions LQR2 and LQR3 do not seem to suffer from any misspecification and perform very well in both tails in-sample. In case of WTI crude oil futures, we observe similar results for the 90\% quantile and some rejection for ARFIMA and CAViaR in the far left tail, although these appear to be marginal at the 5\% level in the latter case. Thus we conclude that the daily semiparametric conditional quantile models perform generally well in-sample, while the ARFIMA-based lognormal-normal mixture appears to be slightly misspecified. Future work might therefore experiment with alternative distributional assumptions in the latter model.\medskip

\noindent \emph{Conditional median}

\noindent The results for the conditional median are substantially different from those for the far left and right tails. This is hardly surprising given the vast body of evidence documenting the lack of predictability of short horizon asset returns. Our estimation results show that the variables we consider have generally little predictive power for the median, either because the estimated coefficient are insignificant or their magnitude is small. The weak evidence for predictability that we find shows that the lagged absolute return and lagged realized measures of volatility are sometimes negatively correlated with future median; see for example the AS and RAS models for S\&P 500 future and the LQR2 and RSAV1 models for WTI Crude Oil futures. This is consistent with the findings of \citeasnoun{bnks10}, and may be due to the leverage effect whereby an increase in volatility maybe followed by a decline in asset prices. The relatively weak statistical significance of our results, however, leads us to believe that a proper test of economic significance needs to be carried out before any definitive conclusions can be drawn; we leave this for future work.

\subsubsection{Out-of-sample performance}
We now assess the out-of-sample performance of the quantile models. We focus on one, five and ten-step-ahead quantile forecasts and adopt the rolling approach, where we keep the estimation window size fixed and forecast the last 500 daily, weekly or 10-day quantiles. The multistep ahead forecasts are obtained from models fitted to the multiperiod returns directly (direct forecasting), except for the ARFIMA-based forecasts, where we use the model fit to the daily time-series to forecast quantiles at all horizons. The parameter estimates from the semiparametric models fitted to the multiperiod returns are not reported to save space, but are available on request. The estimation results for the ARFIMA models are reported in Table \ref{tab:arfima}.

We start by assessing the absolute performance of the one-step-ahead forecasts using the \citeasnoun{bcp11} approach as in the previous section, recalling that this approach is not suitable for multi-step-ahead forecasts. The results are reported in the right-hand side panels of Tables \ref{tab:insampleretSP} and \ref{tab:insampleretOil}. We find that all models perform well. The unconditional coverage is close to the nominal levels and the DQ test signals significant misspecification only in the case of the 90\%-quantile SAV and RSAV1 models for S\&P 500 futures returns. Some minor misspecification is also indicated for the ARFIMA, SAV, RSAV1 and LQR1 models for the median of WTI Crude Oil futures returns.

Turning to the evaluation of relative performance, we report in Tables \ref{tab:outofsampleretSP} and \label{tab:outofsampleretOil} for each $\alpha$-quantile, model and forecast horizon, the out-of-sample unconditional coverage ($\hat{\alpha}$), the value of the tick-loss function given in equation (\ref{eq:tickloss}) and the Diebold-Mariano test statistic for the null hypothesis of equal predictive ability, where the benchmark model throughout is the linear quantile regression model LQR2. Recall that this model includes the lagged continuous and jump variations ($IV$ and $JV$) and the option-implied volatility as regressors. We use it as a benchmark since it belongs to the class of linear quantile regression models with realized measures, which we newly propose and advocate in this paper, and it performs well both in-sample and out-of-sample for $h=1$ in absolute terms as indicated by the $DQ$ test.

Generally, we only find material difference across the competing models for the one-step-ahead forecasts. First, the ARFIMA-based lognormal-normal mixture outperforms the benchmark linear quantile regression LQR2 in the left tail of the distribution, delivering significantly lower tick-loss at the 5\% level despite relatively poorer unconditional coverage. This is the case for both S\&P 500 and WTI Crude Oil futures. A second interesting finding is that the symmetric absolute value CAViaR model of Engle and Manganelli (2004) is beaten by our benchmark linear model both in the left and right tails at the 5\% level in the case of S\&P 500 futures. This is also true for the asymmetric CAViaR model and the 5\% quantile. However, by incorporating lagged realized measure or option-implied volatility restores the performance of the CAViaR model such that it is statistically indistinguishable from our benchmark. In terms of multi-step ahead forecasts, we find small differences between the various models, both for S\&P500 and WTI Crude Oil futures, and no uniform ranking of the models emerges from our exercise.

\subsection{Realized volatility quantiles}
We now turn to modeling and forecasting the quantiles of realized volatility of S\&P 500 futures. We focus on the median and 75\%, 90\% and 95\% quantiles with the latter two being of particular interest to traders or investors exposed to volatility risk. As in the case of returns, we only report estimation results for three different model specifications that we find particularly interesting, noting that a number of alternative model specifications delivering equally accurate quantile forecast can be considered. The results are summarized in Table \ref{tab:qrRVSP}.

We begin by discussing model HARQ1 where we quantile-regress realized volatility on lagged realized volatility, and the average realized volatilities over the past 5 and 22 days. This model is a quantile autoregression of \citeasnoun{kx06} with 22 lags and restricted parameters. We find that all three regressors are highly statistically significant in the models for the median and 75\% quantile, while only $RV_{t,M}^{1/2}$ and $RV_{t,t-5,M}^{1/2}$ remain significant in the models for the far right tail quantiles (90\% and 95\%). The quantiles of realized volatility are therefore less persistent in the right tail of its distribution. Interestingly, the coefficient estimates for $RV_{t,M}^{1/2}$ increase steadily with $\alpha$ thereby capturing the volatility-of-volatility effect observed among others by \citeasnoun{cmpp08} and \citeasnoun{bkpt09}. If the innovations were homoskedastic as in a pure location model, the quantile regression coefficients would be constant (up to estimation error) across all quantiles. We find quite the opposite: in periods of high volatility, the volatility of volatility increases and this pushes a given conditional $\alpha$ quantile further to the right.

\begin{figure}[!h]
\centering
\begin{tabular}{cc}
  \includegraphics[width=75mm,height=68mm]{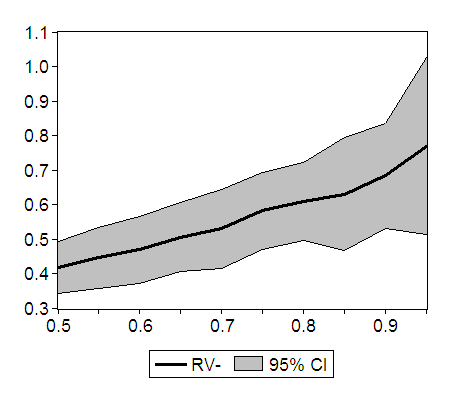} & \includegraphics[width=75mm,height=68mm]{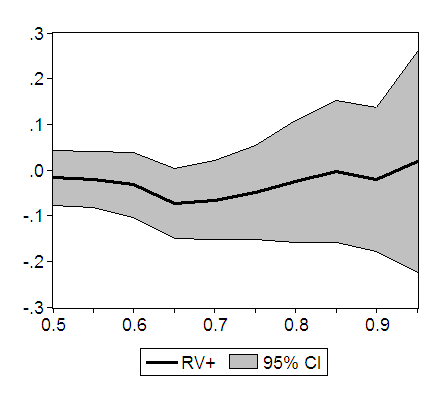} \\
  \includegraphics[width=75mm,height=68mm]{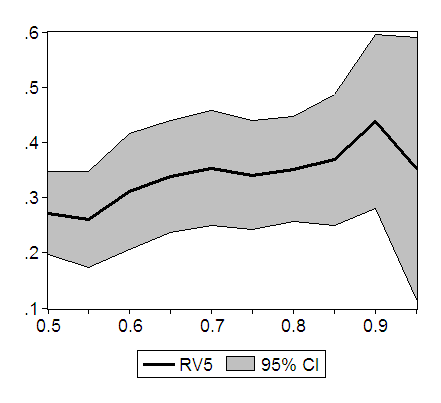} & \includegraphics[width=75mm,height=68mm]{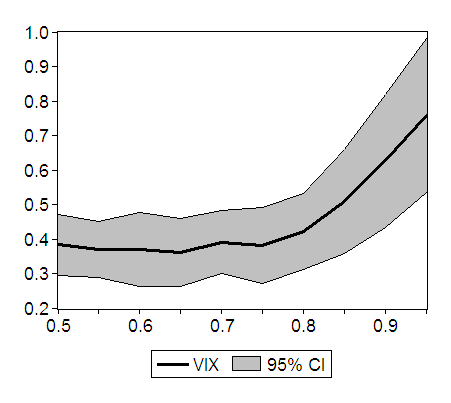}
\end{tabular}
\caption{Estimated Quantile regression process for model HARQ2 in Table \ref{tab:qrRVSP} for S\&P 500 realized volatility. For each $\alpha$-quantile ranging from 0.5 to 0.95, we plot the estimated parameters in the quantile regression $(\widehat{\vbeta}(\alpha))$ together with pointwise 95\% bootstrapped confidence intervals.\label{fig:qrpSPRV}}
\end{figure}

In the HARQ2 model, we augment the set of regressors by implied volatility and replace the lagged realized volatility by upside and downside semi-volatilities. Similarly to the models for daily returns, we find that the downside volatility completely dominates the upside volatility, with the latter being statistically insignificant in all four quantile models (see also Figure \ref{fig:qrpSPRV}). The option-implied volatility possesses significant predictive power for the quantiles of future realized volatility as well and the coefficient estimates increase with $\alpha$ as do the coefficients corresponding to the realized downside semivariance. This implies that the volatility of realized volatility increases not only with historical realized volatility but also with (risk-neutral) expectations of future volatility. Figure \ref{fig:qrpSPRV}  illustrates this effect graphically. The implied volatility also subsumes the effect of $RV_{t,t-22,M}^{1/2}$ in the median and 75\% quantile models. In the models for the 90\% and 95\% quantile, the coefficient estimates on $RV_{t,t-22,M}^{1/2}$ are negative but further investigation reveals that this is due to the presence of insignificant variables in the model; once these are removed all remaining parameter estimates turn out to be positive.

Finally, we study the role of jumps in the quantile models for realized volatility (HARQ3). We find the jump variation variable insignificant on the 5\% level for all quantiles. This result holds irrespective of the presence of implied volatility or $IV_{t,t-22,M}^{1/2}$ in the regressions.

The estimation results for regression quantiles of WTI Crude Oil futures realized volatility are presented in Table \ref{tab:qrRVOil}. Interestingly, we find that the time series of daily realized volatility exhibits a day-of-week pattern: realized volatility tends to be larger on Wednesdays than on other days of the week. This feature is not induced by thin trading associated with holiday periods since these have been removed from our dataset as we mentioned in Section \ref{sec:data}. Nor is it a symptom of price jumps associated with new announcements that are typically made on Wednesdays. The autocorrelation function of the median realized volatility, which is robust to jumps, exhibits the same seasonal pattern as that of the realized volatility. To account for the day-of-week effect, we include a dummy variable, $D_t^{W}$, for Wednesday. As is apparent from Table \ref{tab:qrRVOil}, the Wednesday dummy is statistically significant across all models reported there.

The average realized volatility over the past month, $RV_{t,t-22,M}^{1/2}$, appears to be less important for the prediction of quantiles in the far right tail. Similar decrease in the persistence of conditional quantiles was also observed for the S\&P 500 futures. The difference between the downside and upside realized semivariances in term of predictive power seems to be less pronounced. The coefficient estimates corresponding to $RS^{-}$ are larger than those of $RV^{+}$ but the latter are also marginally statistically significant for 75\% and 90\% quantiles. The jump variation comes out insignificant at conventional levels in all quantile models. Finally, the model-free implied volatility is found to be highly informative for all quantiles of future realized volatility.

Having covered the semiparametric models, we now turn to the fully parametric ARFIMA-based lognormal-normal mixture described in section 5.2. Table \ref{tab:arfima} reports the parameter estimates of ARFIMA(1,d,0) fitted to the time series of logarithmic realized volatilities of S\&P500 and WTI Crude Oil futures contracts. Consistent with previous empirical evidence we find that both series are highly persistent with the long memory parameter $d$ estimated at 0.48 and 0.40, respectively. The first-order autogressive parameter estimates are negative and statistically significant but relatively small.

As in the case of returns, we now assess the absolute in-sample performance of the conditional quantile models using the \citeasnoun{bcp11} test. The results are reported in the left-hand side panel of Table \ref{tab:insampleRV}. Starting with S\&P 500, we find that the log-normal ARFIMA does not fare very well despite having the empirical unconditional coverage close to the nominal level; the $DQ$ test clearly rejects the null hypothesis of correct dynamic specification. The three linear quantile regressions also suffer from some form of dynamic misspecification in case of the median and 75\% quantiles, but exhibit excellent absolute performance in the right tail (90\% and 95\% quantiles). Similar results are obtained for the models for WTI Crude Oil, although here the $DQ$ test indicates misspecification only in the median regressions and at the 10\% significance level.

\subsubsection{Out-of-sample performance}
Finally, we assess the relative out-of-sample performance of the conditional quantiles models for realized volatility. We proceed in the same manner as in the case of returns. We focus on forecasting the last 500 daily, 5-day and 10-day conditional quantiles using the rolling-window approach and direct forecasting, except for the ARFIMA-based forecasts which are based on the ARFIMA model for daily realized volatility and Monte Carlo simulation. For each model, quantile and forecast horizon, we report the unconditional coverage, the value of the tick-loss function and the Diebold-Mariano test statistic for the null hypothesis of equal predictive ability with the benchmark linear quantile regression model HARQ3. We choose this model as benchmark because it performs well in absolute terms in-sample across the different quantiles.

The results are summarized in Table \ref{tab:outofsampleRV}. We find that despite having relatively poor unconditional coverage, the ARFIMA forecasts significantly outperform the linear quantile regressions at the one-day forecast horizon as indicated by the $DM$ test, both the for S\&P500 and WTI Crude Oil. This superior performance, however, disappears at the 5 and 10-day horizons, where the ARFIMA performs on par with the quantile regressions in a statistical sense ($DM$ test), thought the quantile regressions seem to deliver better unconditional coverage and lower value of the tick-loss function for the 90\% and 95\% quantile forecasts, i.e. for the right tail of the realized volatility distribution. Together with the simplicity of the direct forecasting method and the linearity of the model, as opposed to the computationally intensive Monte Carlo, this implies that the linear quantile regressions may be particularly useful in practice for medium-horizon quantile forecasts of realized volatility.

\section{Conclusion}\label{sec:conclusion}
This paper proposes to use linear quantile regression together with realized measures of volatility as covariates to model and forecast conditional quantiles of financial asset returns and realized volatility. Relying on nonparametric measures of the various components of the overall quadratic variation we avoid making restrictive parametric assumptions on the dynamics of the price process. Thanks to the flexibility of quantile regression, we place no assumptions on the distributions of return or volatility innovations, and we are not confined to the class of location-scale models for either returns or realized volatility.

In an empirical application to S\&P 500 futures prices, we document the role of different components of historical volatility as well as option-implied volatility and find that either individually or in a combination deliver accurate in-sample and out-of-sample fit. Applying the methodology to a series of WTI Crude Oil future realized volatility shows that the quantile regression models perform reasonably well even when applied to substantially more volatile and less persistent data. The models can therefore serve as useful risk managements tools for investors trading the futures contracts themselves or various derivative contracts written on realized volatility.

In a comparison with two competing models, the CAViaR of Engle and Manganelli (2004) and the lognormal-normal mixture of Andersen et al. (2003), we find that neither of the models dominate in terms of performance uniformly across different quantiles. Putting realized measures into the CAViaR model does not drive out the other variables in the CAViaR equation completely and it improves its performance. The linear quantile regressions with realized measures, however, seem to perform no worse than the realized CAViaR. The ARFIMA-based lognormal-normal mixture delivers generally poorer unconditional coverage but it often exhibits lower tick-loss at the same time. For medium-horizon realized volatility forecast, we find that the linear quantile regression seems to perform better, especially in the right tail of the distribution. Needles to say, we have not considered all potential competitors for our quantile regressions in this paper, so there may be other models that rely on realized measures and deliver equal or even better quantile forecasts.  We leave a fully-fledged comparison for future work.

\newpage
\bibliographystyle{dcu}
\bibliography{MyBibFull}

\newpage
\appendix
\begin{singlespace}
\small
\section{Proofs}
\noindent \textbf{Proof of Lemma 1.} The first part of the result is proved by \citeasnoun{cds11} for bi-power and tri-power variation of \citeasnoun{b-ns04}. Using the same line of argument as in \citeasnoun{cds11}, one can show that the same result holds for the median realized volatility as well and we omit the proof to save space.

To prove the second result, write
\begin{align*}
|JV_{t,M} - JV_t|  &= |RV_{t,M} - IV_{t,M} - JV_t|, \displaybreak[0] \\
&\leq |RV_{t,M} - IV_t - JV_t| + |IV_{t,M} - IV_t|, \displaybreak[0] \\
&\equiv A_{t,M} + B_{t,M}.
\end{align*}
$B_{t,M}$ was discussed above so we need to focus on $A_{t,M}$.
\begin{align*}
A_{t,M} &= \left|\sum_{i=1}^{M} \left(\int_{t_{i-1}}^{t_i} \sigma_u \mathrm{d}W_u + \sum_{l=1}^{\Delta_i L_t} \kappa_l \right)^2 - \int_{t-1}^t \sigma_u^2 \mathrm{d}u - \sum_{l=1}^{\Delta L_t} \kappa_l^2 \right|, \displaybreak[0] \\
&\leq \left|\sum_{i=1}^{M} \left(\int_{t_{i-1}}^{t_i} \sigma_u \mathrm{d}W_u \right)^2 - \int_{t_{i-1}}^{t_i} \sigma_u^2 \mathrm{d}u \right|
 + \left|2\sum_{i=1}^{M} \left(\int_{t_{i-1}}^{t_i} \sigma_u \mathrm{d}W_u \right)\left(\sum_{l=1}^{\Delta_i L_t} \kappa_l \right) \right| \displaybreak[0] \\
& + \left|\sum_{i=1}^{M}\left[\left(\sum_{l=1}^{\Delta_i L_t} \kappa_l \right)^2 - \sum_{l=1}^{\Delta_i L_t} \kappa_l^2 \right] \right| \displaybreak[0] \\
&\equiv C_{t,M} + D_{t,M} + E_{t,M}.
\end{align*}
Now $C_{t,M}$ is the measurement error associated with realized volatility in the absence of jumps and by \citeasnoun{cds11} we have $\mathbb{E}(|C_{t,M}|^k) = O(M^{-k/2})$. Given Assumptions A2 (existence of moments of jumps) and A3 (finite-activity), we can proceed by assuming that there is at most one jump in every time interval $[t_{i-1},t_i]$. Then $E_{t,M} = 0$ and write $D_{t,M}$ as
\begin{align*}
D_{t,M} &= 2 \sum_{i=1}^M \biggl|\int_{t_{i-1}}^{t_i} \sigma_u \mathrm{d}W_u \biggr| |\kappa_{t_i}| \mathbf{1}_{\{\Delta_i L_t = 1\}} \displaybreak[0] \\
&\leq 2 \sum_{i=1}^M \biggl|\int_{t_{i-1}}^{t_i} \sigma_{t_{i-1}} \mathrm{d}W_u \biggr| |\kappa_{t_i}| \mathbf{1}_{\{\Delta_i L_t = 1\}} + 2 \sum_{i=1}^M \biggl|\int_{t_{i-1}}^{t_i} (\sigma_u - \sigma_{t_{i-1}}) \mathrm{d}W_u \biggr| |\kappa_{t_i}| \mathbf{1}_{\{\Delta_i L_t = 1\}} \displaybreak[0] \\
&= D_{t,M}^{(1)} + D_{t,M}^{(2)}.
\end{align*}
For simplicity we focus on the case of $k=2$ noting that the case of $k>2$ can be treated analogously. Taking expectations,
\begin{align*}
\mathbb{E}[|D_{t,M}^{(1)}|^2] &=  4 \sum_{i_1=1}^M \sum_{i_2=1}^M  \mathbb{E} \left[  \sigma_{t_{i_1 - 1}} \sigma_{t_{i_2 - 1}}  \biggl|\int_{t_{i_1-1}}^{t_{i_1}} \mathrm{d}W_u \biggr| \biggl|\int_{t_{i_2-1}}^{t_{i_2}} \mathrm{d}W_u \biggr| |\kappa_{t_{i_1}}|  |\kappa_{t_{i_2}}|   \mathbf{1}_{\{\Delta_{i_1} L_t = 1\}} \mathbf{1}_{\{\Delta_{i_2} L_t = 1\}}   \right] \displaybreak[0] \\
&= 4 \sum_{i_1=1}^M \sum_{i_2=1}^M \mathbb{E} \left[  \sigma_{t_{i_1 - 1}} \sigma_{t_{i_2 - 1}}  \biggl|\int_{t_{i_1-1}}^{t_{i_1}} \mathrm{d}W_u \biggr| \biggl|\int_{t_{i_2-1}}^{t_{i_2}} \mathrm{d}W_u \biggr| \right] \\
 & \hspace{6cm} \times \mathbb{E}[|\kappa_{t_{i_1}}||\kappa_{t_{i_2}}|] \mathbb{E}[\mathbf{1}_{\{\Delta_{i_1} L_t = 1\}} \mathbf{1}_{\{\Delta_{i_2} L_t = 1\}}]
\end{align*}
By H\"{o}lder inequality, the first expectation is $O(M^{-1})$ provided that $\mathbb{E}[\sigma_u^2] < \infty$, while the second expectation is $O(M^{-2})$ if $i_1 \neq i_2$ and $O(M^{-1})$ if $i_1=i_2$, provided that $\mathbb{E}[\kappa_j^2] < \infty$. Thus, $\mathbb{E}[|D_{t,M}^{(1)}|^2] = O(M^{-1})$. Finally, $\mathbb{E}[|D_{t,M}^{(2)}|^2]$ can not be of higher order than $\mathbb{E}[|D_{t,M}^{(1)}|^2]$, see \citeasnoun{cds11} for details. \hfill $\Box$ \bigskip

\noindent \textbf{Proof of Proposition 1.} To save space, we prove the proposition for $V_t = \{IV_{t,M},JV_{t,M}\}$ and $\mathbf{\beta}_Z = \vzeros$, noting that the others cases can be treated analogously. Simplifying notation we will write $\vbeta = \vbeta(\alpha)$ since $\alpha$ is fixed throughout. Define
\begin{eqnarray*}
z_{t+1}(\mathbf{\vbeta}) &=& QV_{t+1} - \beta_0 - \beta_1 IV_t - \beta_2 JV_t, \\
z_{t+1,M}(\mathbf{\vbeta}) &=& RV_{t+1,M} - \beta_0 - \beta_1 IV_{t,M} - \beta_2 JV_{t,M},
\end{eqnarray*}
and write
\begin{eqnarray*}
&& \vspace{-10mm} |QR_{T,M}(\vbeta) - QR_{T}(\vbeta)| \\
&=& \left|\frac{1}{T}\sum_{t=0}^{T-1} z_{t+1,M}(\vbeta)[\alpha - \mathbf{1}\{z_{t+1,M}(\vbeta) \leq 0\}] - \frac{1}{T}\sum_{t=0}^{T-1} z_{t+1}(\vbeta)[\alpha - \mathbf{1}\{z_{t+1}(\vbeta) \leq 0\}] \right|  \\
&=& \left|\frac{1}{T}\sum_{t=0}^{T-1} (z_{t+1,M}(\vbeta)-z_{t+1}(\vbeta))(\alpha - \mathbf{1}\{z_{t+1,M}(\vbeta) \leq 0\}) - (\mathbf{1}\{z_{t+1,M}(\vbeta) \leq 0\}-\mathbf{1}\{z_{t+1}(\vbeta) \leq 0\})z_{t+1}(\vbeta)\right| \\
&\leq& \frac{1}{T}\sum_{t=0}^{T-1} |z_{t+1,M}(\vbeta) - z_{t+1}(\vbeta)| + \frac{1}{T}\sum_{t=0}^{T-1}|z_{t+1}(\vbeta)||\mathbf{1}\{z_{t+1,M}(\vbeta) \leq 0\}-\mathbf{1}\{z_{t+1}(\beta) \leq 0\}|  \\
&\equiv& A_{T,M} + B_{T,M}.
\end{eqnarray*}
Now
\begin{equation*}
A_{T,M} =   \frac{1}{T} \sum_{t=0}^{T-1} |N_{t,M} - \beta_1 N_{t,M}^{(c)} - \beta_2 N_{t,M}^{(d)} | \leq \frac{1}{T} \sum_{t=1}^{T} |N_{t,M}| + \bar{\beta}_1|N_{t,M}^{(c)}| + \bar{\beta}_2 |N_{t,M}^{(d)}|,
\end{equation*}
where $N_{t,M} = RV_{t,M} - IV_t - JV_t$, $\bar{\beta}_1 = \max |\beta_1|$ and $\bar{\beta}_2 = \max |\beta_2|$, which are well-defined since $\Theta$ is compact. It follows by Markov inequality, stationarity and Lemma 1 that $A_{T,M} = O_p(M^{-1/2})$ uniformly in $\vbeta$.

Turning to $B_{t,M}$, we have
\begin{eqnarray*}
B_{T,M} &=& \frac{1}{T}\sum_{t=0}^{T-1} |z_{t+1}(\vbeta)||\mathbf{1}\{z_{t+1}(\vbeta)+ N_{t,M} - \beta_1 N_{t,M}^{(c)} - \beta_2 N_{t,M}^{(d)} \leq 0\} - \mathbf{1}\{z_{t+1}(\vbeta) \leq 0\}| \\
&\leq& \frac{1}{T}\sum_{t=0}^{T-1} |z_{t+1}(\vbeta)| \mathbf{1}\{- \sup_t |N_{t,M}| - \bar{\beta}_1 \sup_t |N_{t,M}^{(c)}| - \bar{\beta}_2 \sup_t |N_{t,M}^{(d)}| \leq z_{t+1}(\vbeta)\} \\
 && \qquad \qquad \times \mathbf{1}\{z_{t+1}(\vbeta) \leq \sup_t |N_{t,M}| + \bar{\beta}_1 \sup_t |N_{t,M}^{(c)}| + \bar{\beta}_2 \sup_t |N_{t,M}^{(d)}| \}.
\end{eqnarray*}
By Lemma 1, $\mathbb{E}|N_{t,M}^{(d)}|=O(M^{-k/2})$, and hence by Markov inequality
\begin{eqnarray*}
\mathbb{P}\left[\sup_t T^{-\frac{2}{2k-1}}M^{1/2} |N_{t,M}^{(d)}|>\epsilon\right] &\leq& \sum_{t=1}^T \mathbb{P} \left[T^{-\frac{2}{2k-1}}M^{1/2} |N_{t,M}^{(d)}|>\epsilon\right], \\
&\leq& \frac{1}{\epsilon} T^{1-\frac{2k}{2k-1}} M^{k/2} \mathbb{E}[|N_{t,M}^{(d)}|^k], \\
&=& O(T^{1-\frac{2k}{2k-1}}).
\end{eqnarray*}
and similarly for $N_{t,M}^{(c)}$ and $N_{t,M}$. It follows that there exists a constant $c$ such that with probability approaching one as $T,M \rightarrow \infty$
\begin{eqnarray}\label{eq:BtM}
B_{T,M} &\leq& \frac{1}{T}\sum_{t=1}^T |z_{t}(\vbeta)| \mathbf{1}\{ - b c \epsilon T^{\frac{2}{2k-1}}M^{-1/2} \leq z_{t+1}(\vbeta) \leq bc \epsilon T^{\frac{2}{2k-1}}M^{-1/2} \}, \\
&\equiv& B_{t,M}',
\end{eqnarray}
where $b = 1 + \bar{\beta}_1 + \bar{\beta}_2$. Thus, following \citeasnoun{cds11} we can proceed by conditioning on a set on which (\ref{eq:BtM}) holds and focus on $B_{t,M}'$. By Markov and H\"{o}lder inequalities,
\begin{equation*}
\mathbb{P}(B_{T,M}' > \eta) \leq \frac{1}{\eta}\frac{1}{T} \sum_{t=0}^{T-1} \mathbb{E}(z_{t}^2(\vbeta))^{1/2} \mathbb{P}(- b c \epsilon T^{\frac{2}{2k-1}}M^{-1/2} \leq z_{t+1}(\vbeta) \leq bc \epsilon T^{\frac{2}{2k-1}}M^{-1/2})^{1/2}.
\end{equation*}
Since $\Theta$ is compact, Assumption A2 implies that $\mathbb{E}(z_{t}^2(\vbeta))^{1/2}<\infty$ and if $T^{\frac{2}{2k-1}}M^{-1/2} \rightarrow 0$ as $T,M \rightarrow \infty$, we have $\mathbb{P}(- b c \epsilon T^{\frac{2}{2k-1}}M^{-1/2} \leq z_{t+1}(\vbeta) \leq bc \epsilon T^{\frac{2}{2k-1}}M^{-1/2})^{1/2} \rightarrow 0$, uniformly in $\vbeta$. The statement in the proposition then follows.  \hfill $\Box$

\clearpage

\section{Constructing model-free implied volatility for oil}
\subsection{Theoretical Background}
The methodology we employ for constructing model-free implied variance (\emph{ImV}) for oil is based on \citeasnoun{cw09}. The idea of \citeasnoun{cw09} is to synthetise a variance swap contract using European options and futures contracts. Since the oil options are American-style futures options, further adjustments are required to account for the early exercise premium. Here we follow \citeasnoun{ts09}.

To fix ideas, let $F_{t,T}$ denote the time-$t$ futures price of maturity $T > t$, and let $RV_{t,T_1}$, $T_1 \leq T$, denote the realized variance of the futures price between $t$ and $T_1$. A variance swap with notional dollar amount $L$ is a contract that pays at maturity $T_1$ to the long side the following amount
\begin{equation*}
(RV_{t,T_1} - SW_{t,T_1})L.
\end{equation*}
Since the value of the swap at inception is zero, absence of arbitrage requires that
\begin{equation*}
SW_{t,T_1} = \mathrm{E}^{\mathbb{Q}}(RV_{t,T_1}) = ImV_{t,T_1},
\end{equation*}
i.e. the swap rate is equal to the risk-neutral market expectation of future realized variance, that is, the \emph{ImV}. \citeasnoun{cw09} show that the \emph{ImV} can be approximated by
\begin{equation} \label{eq:imv}
ImV_{t,T_1,T} \approx \frac{2}{B_{t,T_1}(T_1 - t)}\left(\int_0^{F_{t,T}}\frac{\mathcal{P}(t,T_1,T,X)}{X^2}\mathrm{d}X + \int_{F_{t,T}}^{\infty} \frac{\mathcal{C}(t,T_1,T,X)}{X^2}\mathrm{d}X \right),
\end{equation}
where $B_{t,T_1}$ is the time-$t$ price of a zero-coupon bond maturing at time $T_1$, and $\mathcal{P}(t,T_1,T,X)$ and $\mathcal{C}(t,T_1,T,X)$ denote the time-$t$ price of a European put and call options, respectively, expiring at time $T_1$ with strike $X$ written a a futures contract with maturity at $T$. When the underlying futures price trajectories are continuous, the relation (\ref{eq:imv}) is exact. In the presence of jumps, a jump error arises but it is shown to be rather small in a simulation exercise by \citeasnoun{cw09}.

To account for the early exercise premium embedded in the American-style options, we resort to the quadratic approximation formulas for American-style puts and call developed by \citeasnoun{baw87} (henceforth BAW). In particular, for each strike and maturity, we first invert the BAW formula to obtain the implied volatility and then plug the implied volatility into \citeasnoun{b76} formula for pricing European-style futures options to obtain $\mathcal{P}(t,T_1,T,X)$ and $\mathcal{C}(t,T_1,T,X)$.

\subsection{Data and implementation}
We use daily settlement prices for WTI Crude Oil futures traded on the New York Mercantile Exchange (NYMEX) and the futures options traded on the Chicago Mercantile Exchange (CME). To proxy for the risk-free interest rates we employ the zero curve supplied by OptionMetrics.  The sample period runs from September 4, 2001 till August 30, 2008.

Before calculating the implied volatility we perform some basic data cleaning. We remove options which have less that 10 days to maturity to avoid possible distortions associated with near-maturity microstructure effects. We also discard all options with prices smaller than 0.05 USD. Finally, we only consider options satisfying the no-arbitrage bounds, see e.g. \citeasnoun{hull00}.

For each day in the sample, we construct the 30-day implied volatility using the two nearest maturities, denoted by $T_1$ and $T_2$, $T_1 < T_2 < T$. For each maturity, we first obtain the implied volatility smile from the available out-of-the money put and call options by inverting the BAW formula. We then linearly interpolate the implied volatilities at different moneyness levels $k = \log(X/F)$. For strikes smaller than the lowest available strike, we use the lowest available strike. Similarly for strikes higher than the highest available one. We thus obtain implied volatilities over a fine grid ranging from -10 to +10 standard deviations from the current futures prices and use the \citeasnoun{b76} formula to convert the implied volatilities into prices of out-of-the-money put and call options. These prices are then used in (\ref{eq:imv}) to approximate $ImV_{t,T_1,T}$ and $ImV_{t,T_2,T}$. Finally, to obtain the 30-day variance swap rate, we linearly interpolate between the two available maturities:
\begin{equation*}
ImV_{t,T^*} = \frac{1}{(T^* - t)} \left[\frac{ImV_{t,T_1,T}(T_1 - t)(T_2 - T^*) + ImV_{t,T_2,T}(T_2 - t)(T^* - T_1)}{T_2 - T_1}\right]
\end{equation*}
where $T^*$ is such that $T^* - t$ is 30 days.
\end{singlespace}

\section*{Tables}
\begin{table}[!h]
\begin{center}
\begin{singlespacing}
\begin{tabular}{lrrr}
\toprule
  &S\&P 500 & & WTI Crude Oil  \\
\cmidrule{2-2}   \cmidrule{4-4}
d & 0.48 & & 0.40 \\
 & (32.3) & & (14.8)  \\
$\phi_1$ & -0.07 & & -0.11 \\
          & (-3.14) & & (-2.82) \\
Mean & -0.37 & & 1.10 \\
$\sigma^2$ & 0.25 & & 0.19 \\
\midrule
Log Lik & -1905.18 & & -726.06 \\
AIC & 1.44 & & 1.17 \\
\bottomrule
\end{tabular}
\caption{ARFIMA parameter estimates with t-statistics in parentheses.\label{tab:arfima}}
\end{singlespacing}
\end{center}
\end{table}

\clearpage

\begin{table}[!h]
\begin{center}
\renewcommand{\baselinestretch}{1.0}
\begin{tabular}{lcccccccc}
\toprule
                    & Mean & Std. Dev. & Skew. & Kurt. & Min & Max & JB & $\mathrm{LB}_{20}$ \\
\cmidrule{2-9}
$r_t$               &  -0.010    &  1.015    &   0.105    &    8.843   &   -7.730    &   8.382   &   1579   &  35.10  \\
$RV_t$              &   0.992    &  1.368    &   8.177    &    129.3   &    0.052    &  32.99    &  19865   &  7166   \\
$RV_t^{1/2}$        &   0.890    &  0.445    &   2.292    &    11.08   &    0.228    &   5.744   &  1673    &  17894  \\
$\log RV_t$         &  -0.434    &  0.881    &   0.283    &    0.109   &    -2.951   &   3.496   &   42.57  &  23807  \\
$RV_t^{-} $         &   0.492    &  0.638    &   5.345    &    46.40   &    0.013    &   9.004   &  15037   &  7896   \\
$RV_t^{+}$          &   0.499    &  0.841    &   13.70    &   329.4    &    0.023    &  26.38    &  69983   &   3647  \\
$MedRV_t$           &   0.919    &  1.248    &   6.228    &   61.24    &    0.042    &  19.13    &  22046   &  9166   \\
$VIX_t$             &   1.278    &  0.834    &   1.621    &    3.657   &    0.267    &   5.731   &   1795   &  48706  \\
\bottomrule
\end{tabular}
\caption{Summary statistics for S\&P 500 futures returns, realized measures and option-implied volatility. JB denotes the Jarque-Bera test statistics for normality and $\mathrm{LB}_{20}$ is the Ljung-Box test statistics for serial correlation up to lag 20. All realized measures are calculated from 5-minute prices obtained from irregularly-spaced transactions data using the last-tick method. The sample period is from January 3, 1997 till June 30, 2008, yielding 3140 daily observations.\label{tab:statssp}}
\end{center}
\end{table}

\begin{table}[!h]
\begin{center}
\renewcommand{\baselinestretch}{1.0}
\begin{tabular}{lcccccccc}
\toprule
                    & Mean & Std. Dev. & Skew. & Kurt. & Min & Max & JB & $\mathrm{LB}_{20}$ \\
\cmidrule{2-9}
$r_t$               &   0.022    &  2.026    &  -0.155    &    3.623   &   -12.53    &   14.43   &   477.9  &  33.02  \\
$RV_t$              &   4.066    &  4.222    &   3.620    &   16.54    &    0.464    &   37.58   &   7942   &  13478  \\
$RV_t^{1/2}$        &   1.734    &  0.545    &   1.689    &   5.802    &    0.681    &    5.573  &   601.9  &   5330  \\
$\log RV_t$         &   1.015    &  0.574    &   0.387    &   0.548    &   -0.766    &    3.435  &    42.06 &   6159  \\
$RV_t^{-} $         &   2.101    &  2.374    &   3.522    &   15.83    &    0.089    &   21.42   &   7266   &  9997   \\
$RV_t^{+}$          &   1.965    &  2.205    &   4.517    &   28.28    &    0.217    &   23.61   &  10004   &   9457  \\
$MedRV_t$           &   3.812    &  3.916    &   3.769    &   19.46    &    0.194    &   43.78   &   7074   &  12641  \\
$ImV_t$             &   1.978    &  0.438    &   1.423    &   2.651    &    1.289    &   4.111   &   794.9  &  23968  \\
\bottomrule
\end{tabular}
\caption{Summary statistics for WTI Crude Oil futures returns, realized measures and option-implied volatility. JB denotes the Jarque-Bera test statistics for normality and $\mathrm{LB}_{20}$ is the Ljung-Box test statistics for serial correlation up to lag 20. All realized measures are calculated from 5-minute prices obtained from irregularly-spaced transactions data using the last-tick method. The sample period is from September 4, 2001  till August 30, 2008, yielding 1870 daily observations.\label{tab:statsoil}}
\end{center}
\end{table}

\clearpage


\begin{sidewaystable}[!h]
\begin{center}
\small
\renewcommand{\tabcolsep}{1.75mm}
\renewcommand{\baselinestretch}{1.0}
\begin{tabular}{lccccccccccccccccc}
\toprule
          & \multicolumn{5}{c}{Linear quantile regressions} & & \multicolumn{5}{c}{CAViaR} & & \multicolumn{5}{c}{Realized CAViaR} \\
            \cmidrule{2-6}  \cmidrule{8-12}  \cmidrule{14-18}
   \multicolumn{1}{r}{$\alpha$} & 0.05 & 0.10 & 0.50 & 0.90 & 0.95 & & 0.05 & 0.10 & 0.50 & 0.90 & 0.95 & & 0.05 & 0.10 & 0.50 & 0.90 & 0.95\\
\midrule
& \multicolumn{5}{c}{LQR1} & & \multicolumn{5}{c}{SAV} & & \multicolumn{5}{c}{RSAV1} \\
           \cmidrule{2-6}  \cmidrule{8-12}  \cmidrule{14-18}
const      & $\underset{(-5.22)}{-0.59}$  & $\underset{(-3.95)}{-0.41}$ &$\underset{(-0.53)}{-0.02}$ &  $\underset{(1.67)}{0.11}$ & $\underset{(2.56)}{0.24}$  & &$\underset{(-7.03)}{-0.06}$ & $\underset{(-8.85)}{-0.05}$ & $\underset{(0.00)}{0.00}$ & $\underset{(5.42)}{0.02}$ & $\underset{(2.60)}{0.04}$ &    &$\underset{(0.19)}{0.00}$ & $\underset{(-0.72)}{-0.01}$ & $\underset{(-0.01)}{-0.00}$ & $\underset{(-0.02)}{-0.00}$ & $\underset{(1.05)}{0.02}$ \\
$q_t$ & & & & & & &$\underset{(34.35)}{0.92}$ & $\underset{(75.45)}{0.91}$ & $\underset{(0.41)}{0.30}$ & $\underset{(119.31)}{0.97}$ & $\underset{(56.73)}{0.96}$ &    &$\underset{(13.80)}{0.80}$ & $\underset{(24.15)}{0.83}$ & $\underset{(0.41)}{0.29}$ & $\underset{(14.63)}{0.77}$ & $\underset{(26.80)}{0.81}$ \\
$|r_t|$ & & & & & & & $\underset{(-3.11)}{-0.17}$ & $\underset{(-7.57)}{-0.15}$ & $\underset{(-2.02)}{-0.07}$ & $\underset{(4.33)}{0.05}$ & $\underset{(3.96)}{0.07}$ &    &$\underset{(0.52)}{0.02}$ & $\underset{(-1.16)}{-0.06}$ & $\underset{(-1.58)}{-0.07}$ & $\underset{(-3.43)}{-0.17}$ & $\underset{(-3.74)}{-0.20}$ \\
$RV_t^{1/2}$    & $\underset{(-9.52)}{-1.12}$ & $\underset{(-7.24)}{-0.88}$ & $\underset{(1.18)}{0.05}$ &  $\underset{(11.9)}{1.15}$ & $\underset{(12.0)}{1.43}$  & & &   &  &   &    &   &$\underset{(-2.52)}{-0.34}$ & $\underset{(-2.14)}{-0.19}$ & $\underset{(0.01)}{0.00}$ & $\underset{(5.76)}{0.37}$ & $\underset{(6.68)}{0.38}$\\
\midrule
& \multicolumn{5}{c}{LQR2} & & \multicolumn{5}{c}{} & & \multicolumn{5}{c}{RSAV2} \\
           \cmidrule{2-6}    \cmidrule{14-18}
const        & $\underset{(-1.18)}{-0.16}$ & $\underset{(-1.17)}{-0.11}$ &$\underset{(-0.93)}{-0.05}$& $\underset{(-2.23)}{-0.19}$ & $\underset{(-2.26)}{-0.17}$  & & & & & & & & $\underset{(2.74)}{0.18}$ & $\underset{(2.21)}{0.14}$ & $\underset{(0.49)}{0.05}$ & $\underset{(-0.13)}{-0.00}$ & $\underset{(0.90)}{0.14}$ \\
$q_t$ & & & & & & & & & & & & & $\underset{(0.43)}{0.07}$ & $\underset{(0.97)}{0.24}$ & $\underset{(-2.22)}{-0.75}$ & $\underset{(9.73)}{0.76}$ & $\underset{(0.71)}{0.18}$\\
$|r_t|$ & & & & & & & & & & & & &$\underset{(0.93)}{0.04}$ & $\underset{(0.28)}{0.02}$ & $\underset{(-0.96)}{-0.01}$ & $\underset{(-3.01)}{-0.16}$ & $\underset{(-2.69)}{-0.23}$ \\
$IV_t^{1/2}$       & $\underset{(-2.09)}{-0.48}$ & $\underset{(-2.60)}{-0.40}$ &$\underset{(0.54)}{0.03}$& $\underset{(3.55)}{0.53}$ & $\underset{(3.96)}{0.67}$  & & & & & & & &$\underset{(-5.13)}{-0.68}$ & $\underset{(-2.86)}{-0.51}$ & $\underset{(1.10)}{0.06}$ & $\underset{(3.82)}{0.36}$ & $\underset{(1.98)}{0.67}$ \\
$JV_t^{1/2}$       & $\underset{(-1.24)}{-0.80}$ & $\underset{(0.11)}{0.10}$ &$\underset{(0.36)}{0.11}$& $\underset{(0.01)}{0.00}$ & $\underset{(0.18)}{0.04}$  & & & & & & & & $\underset{(-0.07)}{-0.03}$ & $\underset{(0.38)}{0.02}$ & $\underset{(-2.01)}{-0.23}$ & $\underset{(0.10)}{0.05}$ & $\underset{(1.22)}{0.80}$ \\
$VIX_t$      & $\underset{(-3.48)}{-0.92}$ & $\underset{(-4.02)}{-0.64}$ &$\underset{(0.60)}{0.05}$& $\underset{(6.07)}{0.79}$ & $\underset{(6.00)}{0.99}$  & & & & & & & & $\underset{(-4.40)}{-0.90}$ & $\underset{(-2.11)}{-0.53}$ & $\underset{(-0.99)}{-0.13}$ & $\underset{(0.39)}{0.02}$ & $\underset{(1.09)}{0.58}$  \\
\midrule
& \multicolumn{5}{c}{LQR3} & & \multicolumn{5}{c}{AS} & & \multicolumn{5}{c}{RAS} \\
           \cmidrule{2-6}  \cmidrule{8-12}  \cmidrule{14-18}
const      & $\underset{(-1.44)}{-0.20}$ & $\underset{(-0.96)}{-0.09}$ &$\underset{(-0.67)}{-0.04}$& $\underset{(-2.42)}{-0.19}$ & $\underset{(-1.62)}{-0.14}$  & & $\underset{(-1.66)}{-0.01}$ & $\underset{(-1.18)}{-0.01}$ & $\underset{(0.75)}{0.01}$ & $\underset{(2.81)}{0.01}$ & $\underset{(1.72)}{0.02}$ &    &$\underset{(1.52)}{0.07}$ & $\underset{(0.29)}{0.01}$ & $\underset{(-0.03)}{-0.00}$ & $\underset{(1.84)}{0.04}$ & $\underset{(1.89)}{0.04}$ \\
$q_t$ & & & & & & & $\underset{(65.44)}{0.94}$ & $\underset{(94.83)}{0.93}$ & $\underset{(6.22)}{0.74}$ & $\underset{(112.36)}{0.97}$ & $\underset{(69.00)}{0.97}$ &    &$\underset{(1.42)}{0.35}$ & $\underset{(16.72)}{0.85}$ & $\underset{(3.25)}{0.70}$ & $\underset{(9.86)}{0.80}$ & $\underset{(17.61)}{0.85}$  \\
  $(r_t)^+$ & & & & & & & $\underset{(-0.12)}{-0.00}$ & $\underset{(-0.60)}{-0.01}$ & $\underset{(2.41)}{0.04}$ & $\underset{(-1.09)}{-0.02}$ & $\underset{(-1.13)}{-0.02}$ &    &$\underset{(1.51)}{0.20}$ & $\underset{(1.40)}{0.07}$ & $\underset{(1.14)}{0.10}$ & $\underset{(-1.70)}{-0.23}$ & $\underset{(-3.01)}{-0.28}$ \\
  $(r_t)^-$  & & & & & & & $\underset{(-7.73)}{-0.20}$ & $\underset{(-10.51)}{-0.19}$ & $\underset{(-4.12)}{-0.09}$ & $\underset{(7.37)}{0.08}$ & $\underset{(4.17)}{0.11}$ &    &$\underset{(0.47)}{0.05}$ & $\underset{(-2.22)}{-0.12}$ & $\underset{(-2.03)}{-0.15}$ & $\underset{(-1.07)}{-0.07}$ & $\underset{(-0.40)}{-0.04}$\\
$RS_t^{{+}^{1/2}}$  & $\underset{(-0.23)}{-0.09}$ & $\underset{(-0.68)}{-0.14}$ &$\underset{(-0.64)}{-0.06}$& $\underset{(-0.02)}{-0.00}$ & $\underset{(1.30)}{0.31}$  & & &   &    &    &    &       &$\underset{(-0.43)}{-0.22}$ & $\underset{(-0.72)}{-0.07}$ & $\underset{(-0.67)}{-0.33}$ & $\underset{(0.73)}{0.36}$ & $\underset{(0.66)}{0.29}$\\
$RS_t^{{-}^{1/2}}$   & $\underset{(-1.51)}{-0.71}$ & $\underset{(-1.91)}{-0.43}$ &$\underset{(1.02)}{0.12}$& $\underset{(3.72)}{0.78}$ & $\underset{(2.08)}{0.73}$  & & &   &    &    &    &       &$\underset{(-1.13)}{-0.69}$ & $\underset{(-0.85)}{-0.13}$ & $\underset{(0.79)}{0.30}$ & $\underset{(0.23)}{0.09}$ & $\underset{(0.27)}{0.13}$ \\
$VIX_t$     & $\underset{(-2.99)}{-0.85}$ & $\underset{(-4.39)}{-0.67}$ &$\underset{(0.33)}{0.03}$& $\underset{(5.27)}{0.78}$ & $\underset{(4.76)}{0.90}$  & & &   &    &    &    &      &$\underset{(-1.71)}{-0.51}$ & $\underset{(-0.96)}{-0.03}$ & $\underset{(0.31)}{0.02}$ & $\underset{(0.27)}{0.02}$ & $\underset{(0.67)}{0.04}$ \\
\bottomrule
\end{tabular}
\caption{Estimated conditional quantile models for daily S\&P500 futures returns (t-statistics in parentheses). The left-hand side panel reports estimation results for the linear quantile regression models with realized measures proposed in this paper. The middle panel reports estimates of the symmetric absolute value (SAV) and asymmetric slope (AS) CAViaR models of Engle and Manganelli (2004). The right-hand side panel reports estimation results for the realized CAViaR models, i.e. CAViAR models augmented by realized measures. The sample periods runs from January 3, 1997 till June 30, 2008.\label{tab:qrsp}}
\end{center}
\end{sidewaystable}

\clearpage


\begin{sidewaystable}[!h]
\begin{center}
\small
\renewcommand{\tabcolsep}{1.75mm}
\renewcommand{\baselinestretch}{1.0}
\begin{tabular}{lccccccccccccccccc}
\toprule
          & \multicolumn{5}{c}{Linear quantile regressions} & & \multicolumn{5}{c}{CAViaR} & & \multicolumn{5}{c}{Realized CAViaR} \\
           \cmidrule{2-6}  \cmidrule{8-12}  \cmidrule{14-18}
    \multicolumn{1}{r}{$\alpha$} & 0.05 & 0.10 & 0.50 & 0.90 & 0.95 & & 0.05 & 0.10 & 0.50 & 0.90 & 0.95 & & 0.05 & 0.10 & 0.50 & 0.90 & 0.95\\
\midrule
& \multicolumn{5}{c}{LQR1} & & \multicolumn{5}{c}{SAV} & & \multicolumn{5}{c}{RSAV1} \\
           \cmidrule{2-6}  \cmidrule{8-12}  \cmidrule{14-18}
const       & $\underset{(-3.55)}{-1.59}$ & $\underset{(-4.74)}{-1.37}$ &$\underset{(0.86)}{0.19}$& $\underset{(6.24)}{1.52}$ & $\underset{(4.23)}{1.87}$  & &$\underset{(-2.18)}{-0.04}$ & $\underset{(-1.50)}{-0.02}$ & $\underset{(-3.03)}{-0.29}$ & $\underset{(1.52)}{0.07}$ & $\underset{(1.21)}{0.06}$ &    &$\underset{(-0.43)}{-0.03}$ & $\underset{(-0.07)}{-0.00}$ & $\underset{(0.17)}{0.01}$ & $\underset{(1.28)}{0.21}$ & $\underset{(1.00)}{0.15}$\\
 $q_t$ & & & & & & & $\underset{(108.83)}{0.97}$ & $\underset{(76.79)}{0.98}$ & $\underset{(-13.87)}{-0.95}$ & $\underset{(26.10)}{0.95}$ & $\underset{(33.79)}{0.96}$ &    &$\underset{(14.85)}{0.90}$ & $\underset{(13.92)}{0.90}$ & $\underset{(4.53)}{0.66}$ & $\underset{(5.06)}{0.77}$ & $\underset{(7.49)}{0.85}$  \\
  $|r_t|$ & & & & & & &$\underset{(-2.90)}{-0.06}$ & $\underset{(-2.66)}{-0.03}$ & $\underset{(0.99)}{0.02}$ & $\underset{(1.49)}{0.04}$ & $\underset{(1.70)}{0.05}$ &    &$\underset{(0.11)}{0.01}$ & $\underset{(1.11)}{0.05}$ & $\underset{(3.04)}{0.10}$ & $\underset{(-0.41)}{-0.02}$ & $\underset{(0.26)}{0.02}$ \\
$RV_t^{1/2}$      & $\underset{(-2.66)}{-0.68}$ & $\underset{(-2.23)}{-0.41}$ &$\underset{(-0.35)}{-0.04}$& $\underset{(2.65)}{0.40}$ & $\underset{(2.31)}{0.59}$  & & &   &  &   &    &   & $\underset{(-1.15)}{-0.15}$ & $\underset{(-1.55)}{-0.15}$ & $\underset{(-2.08)}{-0.09}$ & $\underset{(1.59)}{0.16}$ & $\underset{(1.08)}{0.15}$ \\
\midrule
& \multicolumn{5}{c}{LQR2} & & \multicolumn{5}{c}{} & & \multicolumn{5}{c}{RSAV2} \\
           \cmidrule{2-6}    \cmidrule{14-18}
const       & $\underset{(-1.47)}{-0.68}$ & $\underset{(-0.96)}{-0.40}$ &$\underset{(1.85)}{0.61}$& $\underset{(3.02)}{0.75}$ & $\underset{(1.14)}{0.68}$ & & & & & & & & $\underset{(0.16)}{0.01}$ & $\underset{(0.26)}{0.01}$ & $\underset{(-3.36)}{-0.93}$ & $\underset{(1.40)}{0.42}$ & $\underset{(1.43)}{0.98}$  \\
 $q_t$ & & & & & & & & & & & & & $\underset{(36.30)}{0.95}$ & $\underset{(26.19)}{0.94}$ & $\underset{(-4.54)}{-0.63}$ & $\underset{(-0.06)}{-0.03}$ & $\underset{(-2.92)}{-0.73}$ \\
  $|r_t|$ & & & & & & & & & & & & & $\underset{(-0.14)}{-0.01}$ & $\underset{(1.66)}{0.05}$ & $\underset{(1.57)}{0.04}$ & $\underset{(0.05)}{0.00}$ & $\underset{(-0.89)}{-0.06}$ \\
$IV_t^{1/2}$ &  $\underset{(0.27)}{0.08}$ & $\underset{(0.69)}{0.14}$ &$\underset{(0.95)}{0.17}$& $\underset{(-0.60)}{-0.08}$ & $\underset{(1.06)}{0.30}$ & & & & & & & & $\underset{(-0.72)}{-0.04}$ & $\underset{(-1.75)}{-0.07}$ & $\underset{(-0.75)}{-0.10}$ & $\underset{(-0.80)}{-0.15}$ & $\underset{(1.76)}{0.33}$ \\
$JV_t^{1/2}$ &  $\underset{(-1.46)}{-0.48}$ & $\underset{(-1.27)}{-0.63}$ &$\underset{(0.25)}{0.09}$& $\underset{(0.74)}{0.23}$ & $\underset{(0.49)}{0.29}$ & & & & & & & &$\underset{(-2.66)}{-0.42}$ & $\underset{(-3.06)}{-0.34}$ & $\underset{(-10.35)}{-0.66}$ & $\underset{(1.28)}{0.36}$ & $\underset{(2.32)}{0.34}$\\
$ImV_t$      & $\underset{(-3.60)}{-1.06}$ & $\underset{(-3.82)}{-0.95}$ &$\underset{(-1.94)}{-0.39}$& $\underset{(4.94)}{0.82}$ & $\underset{(2.22)}{0.82}$ & & & & & & & &$\underset{(-1.16)}{-0.03}$ & $\underset{(-0.90)}{-0.02}$ & $\underset{(2.88)}{0.42}$ & $\underset{(1.78)}{1.00}$ & $\underset{(3.34)}{1.68}$\\
\midrule
& \multicolumn{5}{c}{LQR3} & & \multicolumn{5}{c}{AS} & & \multicolumn{5}{c}{RAS} \\
           \cmidrule{2-6}  \cmidrule{8-12}  \cmidrule{14-18}
const       &  $\underset{(-1.06)}{-0.54}$ & $\underset{(-1.28)}{-0.45}$ &$\underset{(2.04)}{0.61}$& $\underset{(2.92)}{0.86}$ & $\underset{(1.01)}{0.62}$ & & $\underset{(-0.60)}{-0.01}$ & $\underset{(-0.49)}{-0.01}$ & $\underset{(-2.78)}{-0.29}$ & $\underset{(1.12)}{0.08}$ & $\underset{(1.54)}{0.15}$ &    &$\underset{(-1.15)}{-0.37}$ & $\underset{(-2.59)}{-0.73}$ & $\underset{(-2.43)}{-0.62}$ & $\underset{(1.08)}{0.57}$ & $\underset{(1.04)}{1.10}$ \\
 $q_t$ & & & & & & &$\underset{(78.54)}{0.96}$ & $\underset{(41.91)}{0.97}$ & $\underset{(-0.86)}{-0.19}$ & $\underset{(16.46)}{0.92}$ & $\underset{(16.43)}{0.90}$ &    &$\underset{(2.98)}{0.56}$ & $\underset{(1.00)}{0.17}$ & $\underset{(-0.80)}{-0.20}$ & $\underset{(-0.15)}{-0.10}$ & $\underset{(-27.22)}{-0.94}$ \\
  $(r_t)^+$ & & & & & & &$\underset{(-1.55)}{-0.05}$ & $\underset{(-0.96)}{-0.03}$ & $\underset{(2.01)}{0.17}$ & $\underset{(0.93)}{0.04}$ & $\underset{(0.71)}{0.05}$ &    &$\underset{(1.13)}{0.18}$ & $\underset{(4.45)}{0.30}$ & $\underset{(2.10)}{0.16}$ & $\underset{(0.00)}{0.00}$ & $\underset{(-0.66)}{-0.04}$\\
  $(r_t)^-$ & & & & & & &$\underset{(-3.02)}{-0.09}$ & $\underset{(-1.84)}{-0.06}$ & $\underset{(-1.28)}{-0.07}$ & $\underset{(1.54)}{0.07}$ & $\underset{(1.88)}{0.15}$ &    &$\underset{(-0.77)}{-0.10}$ & $\underset{(-2.18)}{-0.27}$ & $\underset{(-0.36)}{-0.03}$ & $\underset{(0.36)}{0.06}$ & $\underset{(0.68)}{0.04}$  \\
$RS_t^{{+}^{1/2}}$   & $\underset{(-0.31)}{-0.12}$ & $\underset{(0.13)}{0.03}$ &$\underset{(2.79)}{0.45}$& $\underset{(-1.29)}{-0.35}$ & $\underset{(-0.83)}{-0.23}$  &   &    & &   &    &    &    &$\underset{(-0.13)}{-0.06}$ & $\underset{(-3.81)}{-0.57}$ & $\underset{(-0.12)}{-0.03}$ & $\underset{(0.12)}{0.06}$ & $\underset{(0.99)}{0.20}$ \\
$RS_t^{{-}^{1/2}}$   & $\underset{(-0.02)}{-0.01}$ & $\underset{(0.72)}{0.17}$ &$\underset{(-2.07)}{-0.39}$& $\underset{(1.80)}{0.51}$ & $\underset{(2.12)}{0.58}$  &   &    &  &  &    &    &    &$\underset{(-0.27)}{-0.12}$ & $\underset{(3.36)}{0.72}$ & $\underset{(-0.69)}{-0.15}$ & $\underset{(-0.54)}{-0.28}$ & $\underset{(0.00)}{0.00}$ \\
$ImV_t$       &  $\underset{(-3.57)}{-0.99}$ & $\underset{(-3.82)}{-0.95}$ &$\underset{(-1.38)}{-0.28}$& $\underset{(2.65)}{0.59}$ & $\underset{(2.86)}{0.89}$  &   &    &    & &   &    &    &$\underset{(-1.42)}{-0.36}$ & $\underset{(-3.90)}{-0.69}$ & $\underset{(1.51)}{0.28}$ & $\underset{(1.42)}{0.99}$ & $\underset{(4.09)}{2.07}$ \\
\bottomrule
\end{tabular}
\caption{Estimated conditional quantile models for daily WTI Crude Oil futures returns (t-statistics in parentheses). The left-hand side panel reports estimation results for the linear quantile regression models with realized measures proposed in this paper. The middle panel reports estimates of the symmetric absolute value (SAV) and asymmetric slope (AS) CAViaR models of Engle and Manganelli (2004). The right-hand side panel reports estimation results for the realized CAViaR models, i.e. CAViAR models augmented by realized measures. The sample periods runs from September 4, 2001 until August 30, 2008.\label{tab:qroil}}
\end{center}
\end{sidewaystable}

\clearpage


\begin{table}[!h]
\begin{center}
\small
\renewcommand{\tabcolsep}{1.9mm}
\renewcommand{\baselinestretch}{1.0}
\begin{tabular}{lrrrrrrrrrrrr}
\toprule
  & &\multicolumn{5}{c}{in-sample} & & \multicolumn{5}{c}{out-of-sample}  \\
\cmidrule{3-7}  \cmidrule{9-13}
 & $\alpha$ & 0.05 & 0.10 & 0.50 & 0.90 & 0.95 & & 0.05 & 0.10 & 0.50 & 0.90 & 0.95  \\
\cmidrule{3-7}  \cmidrule{9-13}
ARFIMA & $\hat{\alpha}$ &0.050 & 0.102 & 0.482 & 0.880 & 0.935 && 0.066 & 0.116 & 0.478 & 0.904 & 0.950 \\
 & $DQ$ & 1.831 & 2.177 & 16.92 & 27.26 & 16.53 && 7.829 & 7.745 & 4.274 & 12.89 & 8.488 \\
 & $p$-val & 0.939 & 0.904 & 0.013 & 0.000 & 0.023 && 0.336 & 0.287 & 0.675 & 0.054 & 0.257 \\
\cmidrule{3-7}  \cmidrule{9-13}
SAV& $\hat{\alpha}$  &0.050 & 0.100 & 0.501 & 0.900 & 0.950 && 0.050 & 0.092 & 0.506 & 0.926 & 0.966 \\
  & $DQ$& 5.217 & 8.195 & 12.14 & 20.39 & 16.16 && 5.789 & 7.104 & 4.169 & 22.94 & 7.006 \\
 & $p$-val  & 0.536 & 0.237 & 0.063 & 0.003 & 0.025 && 0.579 & 0.309 & 0.676 & 0.001 & 0.433 \\
\cmidrule{3-7}  \cmidrule{9-13}
RSAV1& $\hat{\alpha}$  &0.050 & 0.100 & 0.501 & 0.901 & 0.951 && 0.074 & 0.118 & 0.508 & 0.896 & 0.948 \\
 & $DQ$ & 2.048 & 3.748 & 10.31 & 18.61 & 3.658 &&  6.113 & 2.581 & 5.014 & 15.59 & 10.25 \\
 & $p$-val  & 0.901 & 0.717 & 0.115 & 0.007 & 0.725 && 0.540 & 0.864 & 0.568 & 0.027 & 0.122  \\
\cmidrule{3-7}  \cmidrule{9-13}
RSAV2& $\hat{\alpha}$  &0.050 & 0.100 & 0.501 & 0.901 & 0.952 && 0.070 & 0.116 & 0.510 & 0.896 & 0.948 \\
 & $DQ$ & 6.033 & 3.762 & 11.03 & 18.43 & 3.225 && 5.805 & 2.711 & 3.181 & 6.246 & 9.235 \\
 & $p$-val  & 0.436 & 0.699 & 0.096 & 0.007 & 0.773 && 0.581 & 0.868 & 0.800 & 0.429 & 0.182 \\
\cmidrule{3-7}  \cmidrule{9-13}
AS & $\hat{\alpha}$ &0.050 & 0.100 & 0.502 & 0.900 & 0.949 && 0.054 & 0.094 & 0.488 & 0.910 & 0.952 \\
 & $DQ$ & 6.493 & 2.011 & 1.266 & 10.81 & 8.185 && 6.409 & 4.768 & 1.538 & 5.365 & 7.915 \\
 & $p$-val  & 0.402 & 0.911 & 0.970 & 0.103 & 0.239 && 0.512 & 0.601 & 0.951 & 0.517 & 0.323 \\
\cmidrule{3-7}  \cmidrule{9-13}
RAS& $\hat{\alpha}$  &0.050 & 0.101 & 0.501 & 0.900 & 0.951 & &0.054 & 0.102 & 0.496 & 0.908 & 0.958 \\
 & $DQ$ & 5.658 & 0.948 & 0.313 & 8.482 & 0.903 && 1.192 & 2.202 & 1.326 & 12.49 & 6.290 \\
 & $p$-val  & 0.463 & 0.990 & 1.000 & 0.225 & 0.989 && 0.987 & 0.902 & 0.975 & 0.058 & 0.545 \\
\cmidrule{3-7}  \cmidrule{9-13}
LQR1& $\hat{\alpha}$ &0.050 & 0.100 & 0.500 & 0.900 & 0.950 && 0.056 & 0.096 & 0.496 & 0.918 & 0.964 \\
 & $DQ$ & 6.183 & 3.063 & 14.43 & 15.14 & 8.434 && 4.515 & 6.179 & 3.225 & 9.359 & 7.750 \\
 & $p$-val  & 0.398 & 0.802 & 0.025 & 0.023 & 0.226 && 0.760 & 0.399 & 0.780 & 0.191 & 0.331 \\
\cmidrule{3-7}  \cmidrule{9-13}
LQR2& $\hat{\alpha}$ &0.050 & 0.100 & 0.500 & 0.900 & 0.950 && 0.056 & 0.104 & 0.490 & 0.902 & 0.954 \\
 & $DQ$ & 2.067 & 1.092 & 13.09 & 8.941 & 3.918 && 4.581 & 4.357 & 3.611 & 10.67 & 7.513 \\
 & $p$-val  & 0.929 & 0.983 & 0.037 & 0.176 & 0.681 && 0.744 & 0.616 & 0.721 & 0.126 & 0.393 \\
\cmidrule{3-7}  \cmidrule{9-13}
LQR3& $\hat{\alpha}$ &0.050 & 0.100 & 0.500 & 0.900 & 0.950 && 0.052 & 0.104 & 0.494 & 0.898 & 0.956 \\
 & $DQ$ & 4.688 & 1.255 & 12.13 & 10.31 & 3.497 && 3.578 & 7.499 & 3.319 & 6.723 & 6.576 \\
 & $p$-val  & 0.591 & 0.975 & 0.050 & 0.109 & 0.771 && 0.862 & 0.313 & 0.761 & 0.387 & 0.503 \\
\bottomrule
\end{tabular}
\caption{Absolute performance of alternative conditional quantile models for daily S\&P500 futures returns. The left-hand side panel reports results for in-sample performance and the right-hand side panel reports results for out-of-sample performance (one-step-ahead forecasts). For each model and quantile ($\alpha$) we report the unconditional coverage ($\hat{\alpha}$), the Berkowitz et al. (2011) test statistic for correct dynamic specification ($DQ$) and the corresponding Monte Carlo-based p-value ($p$-val).\label{tab:insampleretSP}}
\end{center}
\end{table}

\clearpage


\begin{table}[!h]
\begin{center}
\small
\renewcommand{\tabcolsep}{1.9mm}
\renewcommand{\baselinestretch}{1.0}
\begin{tabular}{lrrrrrrrrrrrr}
\toprule
  & &\multicolumn{5}{c}{in-sample} & & \multicolumn{5}{c}{out-of-sample} \\
\cmidrule{3-7}  \cmidrule{9-13}
 & $\alpha$ & 0.05 & 0.10 & 0.50 & 0.90 & 0.95 & & 0.05 & 0.10 & 0.50 & 0.90 & 0.95 \\
\cmidrule{3-7}  \cmidrule{9-13}

ARFIMA & $\hat{\alpha}$ & 0.030 & 0.075 & 0.485 & 0.898 & 0.951 && 0.050 & 0.102 & 0.484 & 0.871 & 0.942  \\
 & $DQ$ & 19.01 & 12.52 & 9.882 & 10.89 & 7.487 && 5.606 & 5.650 & 14.06 & 9.599 & 7.711 \\
 & $p$-val &  0.005 & 0.040 & 0.137 & 0.076 & 0.339 &&  0.638 & 0.488 & 0.035 & 0.160 & 0.356 \\
\cmidrule{3-7}  \cmidrule{9-13}
SAV& $\hat{\alpha}$  & 0.048 & 0.099 & 0.500 & 0.900 & 0.951 && 0.050 & 0.098 & 0.506 & 0.880 & 0.932 \\
  & $DQ$& 13.84 & 4.236 & 6.056 & 14.13 & 4.002 && 5.406 & 9.114 & 13.92 & 7.815 & 7.993 \\
 & $p$-val  & 0.049 & 0.635 & 0.419 & 0.031 & 0.709 && 0.662 & 0.181 & 0.043 & 0.298 & 0.303 \\
\cmidrule{3-7}  \cmidrule{9-13}
RSAV1& $\hat{\alpha}$  & 0.049 & 0.099 & 0.502 & 0.901 & 0.952 && 0.048 & 0.112 & 0.490 & 0.861 & 0.924 \\
 & $DQ$ & 14.27 & 5.840 & 6.684 & 13.71 & 5.441 && 5.001 & 6.057 & 13.65 & 10.64 & 11.14 \\
 & $p$-val  & 0.045 & 0.464 & 0.364 & 0.032 & 0.545  && 0.720 & 0.449 & 0.037 & 0.118 & 0.091 \\
\cmidrule{3-7}  \cmidrule{9-13}
RSAV2& $\hat{\alpha}$  & 0.052 & 0.099 & 0.500 & 0.899 & 0.951 && 0.048 & 0.116 & 0.508 & 0.871 & 0.940 \\
 & $DQ$ & 6.951 & 7.863 & 5.089 & 17.46 & 8.358  && 5.001 & 10.05 & 6.583 & 8.673 & 4.680 \\
 & $p$-val  & 0.397 & 0.239 & 0.542 & 0.009 & 0.301 && 0.725 & 0.140 & 0.370 & 0.203 & 0.744 \\
\cmidrule{3-7}  \cmidrule{9-13}
AS & $\hat{\alpha}$ & 0.049 & 0.099 & 0.499 & 0.902 & 0.952 && 0.048 & 0.100 & 0.512 & 0.876 & 0.938 \\
 & $DQ$ & 14.27 & 2.081 & 2.311 & 13.21 & 7.755  && 5.001 & 5.334 & 9.258 & 8.169 & 4.529 \\
 & $p$-val  & 0.043 & 0.914 & 0.900 & 0.055 & 0.308 && 0.711 & 0.507 & 0.164 & 0.246 & 0.747 \\
\cmidrule{3-7}  \cmidrule{9-13}
RAS& $\hat{\alpha}$  & 0.051 & 0.099 & 0.501 & 0.901 & 0.949 && 0.050 & 0.102 & 0.506 & 0.876 & 0.932 \\
 & $DQ$ & 6.545 & 10.98 & 1.975 & 5.518 & 5.394 && 4.937 & 3.300 & 7.547 & 5.625 & 7.589 \\
 & $p$-val  & 0.402 & 0.081 & 0.944 & 0.470 & 0.521 &&  0.694 & 0.792 & 0.253 & 0.494 & 0.338 \\
\cmidrule{3-7}  \cmidrule{9-13}
LQR1& $\hat{\alpha}$ &0.051 & 0.100 & 0.500 & 0.900 & 0.949 && 0.048 & 0.098 & 0.514 & 0.900 & 0.952 \\
 & $DQ$ & 3.853 & 2.197 & 18.79 & 12.70 & 3.935 && 2.586 & 6.070 & 11.60 & 3.277 & 6.096 \\
 & $p$-val  & 0.723 & 0.905 & 0.008 & 0.049 & 0.700 && 0.932 & 0.469 & 0.068 & 0.812 & 0.568 \\
\cmidrule{3-7}  \cmidrule{9-13}
LQR2& $\hat{\alpha}$ & 0.051 & 0.100 & 0.500 & 0.901 & 0.949  && 0.046 & 0.112 & 0.514 & 0.892 & 0.948 \\
 & $DQ$ & 1.636 & 3.476 & 14.87 & 9.567 & 3.608 && 5.072 & 7.817 & 10.56 & 2.585 & 4.801 \\
 & $p$-val  & 0.952 & 0.745 & 0.024 & 0.151 & 0.740 && 0.703 & 0.279 & 0.104 & 0.871 & 0.762 \\
\cmidrule{3-7}  \cmidrule{9-13}
LQR3& $\hat{\alpha}$ &0.051 & 0.100 & 0.500 & 0.899 & 0.949 && 0.046 & 0.116 & 0.508 & 0.898 & 0.950 \\
 & $DQ$ & 2.355 & 2.747 & 4.590 & 3.950 & 3.852  && 5.072 & 8.047 & 7.848 & 4.130 & 7.915 \\
 & $p$-val  & 0.887 & 0.850 & 0.621 & 0.684 & 0.715 && 0.678 & 0.235 & 0.230 & 0.674 & 0.305 \\
\bottomrule
\end{tabular}
\caption{Absolute performance of alternative conditional quantile models for daily WTI Crude Oil futures returns. The left-hand side panel reports results for in-sample performance and the right-hand side panel reports results for out-of-sample performance (one-step-ahead forecasts). For each model and quantile ($\alpha$) we report the unconditional coverage ($\hat{\alpha}$), the Berkowitz et al. (2011) test statistic for correct dynamic specification ($DQ$) and the corresponding Monte Carlo-based p-value ($p$-val).\label{tab:insampleretOil}}
\end{center}
\end{table}


\begin{sidewaystable}[!h]
\begin{center}
\footnotesize
\renewcommand{\baselinestretch}{1.0}
\renewcommand{\tabcolsep}{1.5mm}
\begin{tabular}{lrrrrrrrrrrrrrrrrrr}
\toprule
  & &\multicolumn{5}{c}{h=1} & & \multicolumn{5}{c}{h=5} & & \multicolumn{5}{c}{h=10} \\
\cmidrule{3-7}  \cmidrule{9-13} \cmidrule{15-19}
 & $\alpha$ & 0.05 & 0.10 & 0.50 & 0.90 & 0.95 & & 0.05 & 0.10 & 0.50 & 0.90 & 0.95 & & 0.05 & 0.10 & 0.50 & 0.90 & 0.95 \\
\cmidrule{3-7}  \cmidrule{9-13} \cmidrule{15-19}

ARFIMA & $\hat{\alpha}$ &  0.066 & 0.116 & 0.478 & 0.904 & 0.950 & & 0.062 & 0.108 & 0.476 & 0.916 & 0.970 & & 0.064 & 0.092 & 0.452 & 0.962 & 0.984 \\
 & $\hat{L}$ & 0.088 & 0.149 & 0.318 & 0.138 & 0.082 & & 0.200 & 0.319 & 0.624 & 0.264 & 0.154 & & 0.264 & 0.426 & 0.817 & 0.362 & 0.225 \\
 & $DM$ & -4.608\Cross  & -4.819\Cross  & -0.769  & 0.593  & -0.089  & & 0.297  & -0.148  & -1.344 & -0.098  & 1.267  & & 0.976  & 0.324  & -1.686\Cross  & 0.891  & 0.787  \\
\cmidrule{3-7}  \cmidrule{9-13} \cmidrule{15-19}
SAV& $\hat{\alpha}$ &0.050 & 0.092 & 0.506 & 0.926 & 0.966 & & 0.046 & 0.094 & 0.496 & 0.924 & 0.966 & & 0.034 & 0.078 & 0.448 & 0.950 & 0.972 \\
 & $\hat{L}$ & 0.103 & 0.164 & 0.318 & 0.144 & 0.091 & & 0.200 & 0.330 & 0.627 & 0.272 & 0.162 & & 0.254 & 0.428 & 0.825 & 0.353 & 0.222 \\
 & $DM$ & 2.514* & 1.990* & -0.626  & 2.176* & 2.733* & & 0.373  & 1.290 & -0.869  & 0.937  & 2.793* & & 0.526  & 0.769  & -1.495 & 0.306  & 0.738  \\
\cmidrule{3-7}  \cmidrule{9-13} \cmidrule{15-19}
RSAV1& $\hat{\alpha}$ &0.074 & 0.118 & 0.508 & 0.896 & 0.948 & & 0.052 & 0.084 & 0.490 & 0.888 & 0.958 & & 0.056 & 0.092 & 0.434 & 0.910 & 0.962 \\
 & $\hat{L}$ & 0.098 & 0.159 & 0.318 & 0.141 & 0.084 & & 0.203 & 0.324 & 0.632 & 0.266 & 0.153 & & 0.280 & 0.435 & 0.837 & 0.361 & 0.222 \\
 & $DM$ & 0.796  & -0.293  & -0.304  & 2.047* & 1.184  & & 0.922  & 0.471  & -0.018  & 0.351  & 1.510 & & 1.837* & 1.005  & -0.789  & 1.047  & 1.262  \\
\cmidrule{3-7}  \cmidrule{9-13} \cmidrule{15-19}
RSAV2& $\hat{\alpha}$ &0.070 & 0.116 & 0.510 & 0.896 & 0.948 & & 0.052 & 0.090 & 0.490 & 0.882 & 0.950 & & 0.054 & 0.082 & 0.454 & 0.916 & 0.960 \\
 & $\hat{L}$ & 0.097 & 0.159 & 0.318 & 0.139 & 0.082 & & 0.201 & 0.321 & 0.633 & 0.268 & 0.152 & & 0.267 & 0.410 & 0.827 & 0.359 & 0.223 \\
 & $DM$ & 0.590  & -0.151  & -0.279  & 3.224* & 0.792  & & 0.525  & 0.054  & 0.224  & 0.855  & 1.431 & & 1.436 & -1.383 & -1.434 & 1.047  & 1.579 \\
\cmidrule{3-7}  \cmidrule{9-13} \cmidrule{15-19}
AS & $\hat{\alpha}$&0.054 & 0.094 & 0.488 & 0.910 & 0.952 & & 0.048 & 0.088 & 0.492 & 0.928 & 0.960 & & 0.036 & 0.074 & 0.456 & 0.932 & 0.976 \\
 & $\hat{L}$ & 0.101 & 0.161 & 0.316 & 0.139 & 0.084 & & 0.195 & 0.326 & 0.622 & 0.260 & 0.152 & & 0.259 & 0.428 & 0.843 & 0.347 & 0.207 \\
 & $DM$ & 1.965* & 0.600  & -1.589 & 0.902  & 1.163  & & -0.310  & 0.687  & -1.595 & -0.509  & 0.731  & & 1.239  & 0.666  & 0.167  & 0.016  & -0.280  \\
\cmidrule{3-7}  \cmidrule{9-13} \cmidrule{15-19}
RAS & $\hat{\alpha}$&0.054 & 0.102 & 0.496 & 0.908 & 0.958 & & 0.050 & 0.082 & 0.494 & 0.884 & 0.946 & & 0.048 & 0.086 & 0.456 & 0.908 & 0.956 \\
 & $\hat{L}$ & 0.098 & 0.157 & 0.316 & 0.135 & 0.083 & & 0.200 & 0.324 & 0.629 & 0.267 & 0.151 & & 0.265 & 0.415 & 0.851 & 0.363 & 0.218 \\
 & $DM$ & 0.814  & -1.028  & -1.659\Cross & -0.877  & 0.425  & & 0.516  & 0.536  & -0.581  & 0.757  & 1.112  & & 1.730* & -0.644  & 0.748  & 1.037  & 0.386  \\
\cmidrule{3-7}  \cmidrule{9-13} \cmidrule{15-19}
LQR1 & $\hat{\alpha}$&0.056 & 0.096 & 0.496 & 0.918 & 0.964 & & 0.050 & 0.078 & 0.484 & 0.900 & 0.958 & & 0.046 & 0.072 & 0.436 & 0.928 & 0.970 \\
 & $\hat{L}$ & 0.098 & 0.161 & 0.317 & 0.137 & 0.083 & & 0.203 & 0.329 & 0.631 & 0.262 & 0.152 & & 0.266 & 0.434 & 0.838 & 0.347 & 0.210 \\
 & $DM$ & 1.472 & 1.111  & -1.506 & 0.299  & 0.534  & & 1.029  & 1.457 & -0.897  & -0.357  & 1.022  & & 2.363* & 2.116* & -0.886  & -0.094  & -0.281  \\
\cmidrule{3-7}  \cmidrule{9-13} \cmidrule{15-19}
LQR3 & $\hat{\alpha}$&0.052 & 0.104 & 0.494 & 0.898 & 0.956 & & 0.048 & 0.082 & 0.480 & 0.880 & 0.948 & & 0.044 & 0.084 & 0.430 & 0.910 & 0.960 \\
 & $\hat{L}$ & 0.099 & 0.159 & 0.318 & 0.136 & 0.081 & & 0.197 & 0.320 & 0.631 & 0.260 & 0.146 & & 0.249 & 0.420 & 0.841 & 0.349 & 0.212 \\
 & $DM$ & 1.371 & 0.224  & -0.747  & -1.293 & -0.769  & & 0.244  & -0.161  & -0.915  & -2.093\Cross & -0.616  & & 0.223  & -0.487  & -0.651  & 0.687  & -0.722  \\
\cmidrule{3-7}  \cmidrule{9-13} \cmidrule{15-19}
LQR2 & $\hat{\alpha}$&0.056 & 0.104 & 0.490 & 0.902 & 0.954 & & 0.046 & 0.080 & 0.478 & 0.878 & 0.950 & & 0.042 & 0.084 & 0.432 & 0.908 & 0.962 \\
 & $\hat{L}$ & 0.096 & 0.159 & 0.318 & 0.137 & 0.082 & & 0.197 & 0.320 & 0.632 & 0.264 & 0.147 & & 0.249 & 0.421 & 0.842 & 0.348 & 0.213 \\
 \bottomrule
\end{tabular}
\caption{Relative performance of alternative out-of-sample forecasts of S\&P 500 futures return quantiles. For each model, quantile ($\alpha$) and forecasts horizon ($h$), we report the unconditional coverage ($\hat{\alpha}$), the value of the tick-loss function ($\hat{L}$) and the Diebold-Mariano test statistic for equal predictive accuracy with the linear quantile regression model LQR2 serving as the benchmark. We use $*$ to denote significantly less accurate forecasts and \Cross  to denote significantly more accurate forecasts with respect to the benchmark at the 5\% significance level.\label{tab:outofsampleretSP}}
\end{center}
\end{sidewaystable}

\clearpage


\begin{sidewaystable}[!h]
\begin{center}
\footnotesize
\renewcommand{\baselinestretch}{1.0}
\renewcommand{\tabcolsep}{1.5mm}
\begin{tabular}{lrrrrrrrrrrrrrrrrrr}
\toprule
  & &\multicolumn{5}{c}{h=1} & & \multicolumn{5}{c}{h=5} & & \multicolumn{5}{c}{h=10} \\
\cmidrule{3-7}  \cmidrule{9-13} \cmidrule{15-19}
 & $\alpha$ & 0.05 & 0.10 & 0.50 & 0.90 & 0.95 & & 0.05 & 0.10 & 0.50 & 0.90 & 0.95 & & 0.05 & 0.10 & 0.50 & 0.90 & 0.95 \\
\cmidrule{3-7}  \cmidrule{9-13} \cmidrule{15-19}

ARFIMA & $\hat{\alpha}$& 0.050 & 0.102 & 0.484 & 0.871 & 0.942 & & 0.048 & 0.094 & 0.458 & 0.863 & 0.942 & & 0.068 & 0.104 & 0.434 & 0.857 & 0.930 \\
 & $\hat{L}$ &0.162 & 0.277 & 0.647 & 0.286 & 0.167 & & 0.372 & 0.630 & 1.439 & 0.578 & 0.321 & & 0.602 & 0.999 & 2.154 & 0.841 & 0.463 \\
 & $DM$ & -3.188\Cross & -2.741\Cross & -0.429  & 0.027  & -0.427  & & -1.215  & -1.410 & 1.320 & 1.015  & -0.637  & & 0.532  & 0.113  & 0.932  & 1.215  & 1.294 \\
 \cmidrule{3-7}  \cmidrule{9-13} \cmidrule{15-19}
SAV & $\hat{\alpha}$&0.050 & 0.098 & 0.506 & 0.880 & 0.932 & & 0.038 & 0.108 & 0.494 & 0.916 & 0.956 & & 0.054 & 0.102 & 0.504 & 0.896 & 0.954 \\
 & $\hat{L}$ & 0.178 & 0.292 & 0.650 & 0.291 & 0.173 & & 0.380 & 0.638 & 1.427 & 0.557 & 0.324 & & 0.590 & 1.009 & 2.125 & 0.814 & 0.451 \\
 & $DM$ & 1.616 & 0.857  & 0.094  & 1.228  & 1.209  & & -1.126  & -0.470  & 1.120  & -0.820  & -0.291  & & 0.176  & 0.618  & 0.812  & 0.383  & 0.086  \\
\cmidrule{3-7}  \cmidrule{9-13} \cmidrule{15-19}
RSAV1 & $\hat{\alpha}$&0.048 & 0.112 & 0.490 & 0.861 & 0.924 & & 0.040 & 0.104 & 0.470 & 0.863 & 0.922 & & 0.056 & 0.108 & 0.426 & 0.819 & 0.926 \\
 & $\hat{L}$ & 0.171 & 0.287 & 0.651 & 0.290 & 0.171 & & 0.374 & 0.645 & 1.444 & 0.573 & 0.327 & & 0.602 & 1.036 & 2.188 & 0.831 & 0.470 \\
 & $DM$ & 0.071  & -0.115  & 0.339  & 0.887  & 0.499  & & -1.736\Cross & -0.143  & 1.137  & 0.494  & 0.112  & & 0.555  & 0.886  & 1.015  & 0.703  & 0.786  \\
\cmidrule{3-7}  \cmidrule{9-13} \cmidrule{15-19}
RSAV2 & $\hat{\alpha}$&0.048 & 0.116 & 0.508 & 0.871 & 0.940 & & 0.066 & 0.131 & 0.464 & 0.865 & 0.920 & & 0.070 & 0.100 & 0.448 & 0.833 & 0.922 \\
 & $\hat{L}$ & 0.172 & 0.290 & 0.650 & 0.290 & 0.167 & & 0.400 & 0.697 & 1.391 & 0.568 & 0.328 & & 0.647 & 1.069 & 2.047 & 0.855 & 0.482 \\
 & $DM$ & 0.674  & 1.399 & 0.227  & 0.916  & -0.360  & & 0.527  & 1.333 & -0.568  & 0.260  & 0.179  & & 0.806  & 0.918  & -0.511  & 1.058  & 1.035  \\
\cmidrule{3-7}  \cmidrule{9-13} \cmidrule{15-19}
AS & $\hat{\alpha}$&0.048 & 0.100 & 0.512 & 0.876 & 0.938 & & 0.052 & 0.112 & 0.506 & 0.914 & 0.952 & & 0.072 & 0.127 & 0.508 & 0.902 & 0.958 \\
 & $\hat{L}$ & 0.175 & 0.289 & 0.644 & 0.293 & 0.173 & & 0.391 & 0.665 & 1.421 & 0.568 & 0.326 & & 0.637 & 1.070 & 2.146 & 0.817 & 0.450 \\
 & $DM$ & 0.977  & 0.446  & -0.827  & 1.405 & 1.054  & & 0.224  & 0.953  & 0.848  & 0.589  & 0.017  & & 1.472 & 2.447* & 1.298 & 0.419  & 0.057  \\
\cmidrule{3-7}  \cmidrule{9-13} \cmidrule{15-19}
RAS1& $\hat{\alpha}$ &0.050 & 0.102 & 0.506 & 0.876 & 0.932 & & 0.042 & 0.094 & 0.438 & 0.867 & 0.918 & & 0.050 & 0.094 & 0.418 & 0.825 & 0.904 \\
 & $\hat{L}$ & 0.175 & 0.289 & 0.643 & 0.290 & 0.166 & & 0.390 & 0.668 & 1.415 & 0.599 & 0.339 & & 0.581 & 1.004 & 2.098 & 0.863 & 0.502 \\
 & $DM$ & 1.913* & 0.467  & -1.080  & 0.795  & -0.818  & & 0.010  & 0.923  & 0.180  & 1.139  & 0.608  & & -0.157  & 0.153  & 0.058  & 1.129  & 1.304 \\
\cmidrule{3-7}  \cmidrule{9-13} \cmidrule{15-19}
LQR1& $\hat{\alpha}$&0.048 & 0.098 & 0.514 & 0.900 & 0.952 & & 0.042 & 0.106 & 0.500 & 0.916 & 0.952 & & 0.042 & 0.100 & 0.504 & 0.900 & 0.958 \\
 & $\hat{L}$ & 0.171 & 0.288 & 0.650 & 0.287 & 0.173 & & 0.378 & 0.638 & 1.436 & 0.563 & 0.322 & & 0.600 & 1.010 & 2.147 & 0.805 & 0.449 \\
 & $DM$ & 0.238  & 0.204  & -0.429  & 0.128  & 0.818  & & -1.250  & -0.883  & 2.005* & -0.364  & -0.910  & & 0.507  & 0.617  & 2.023* & 0.020  & -0.086  \\
\cmidrule{3-7}  \cmidrule{9-13} \cmidrule{15-19}
LQR3& $\hat{\alpha}$&0.046 & 0.116 & 0.508 & 0.898 & 0.950 & & 0.044 & 0.108 & 0.492 & 0.900 & 0.950 & & 0.050 & 0.102 & 0.490 & 0.892 & 0.950 \\
 & $\hat{L}$ & 0.173 & 0.288 & 0.645 & 0.286 & 0.170 & & 0.383 & 0.650 & 1.412 & 0.563 & 0.322 & & 0.588 & 0.997 & 2.084 & 0.799 & 0.443 \\
 & $DM$ & 1.738* & 0.464  & -1.241  & -0.095  & 0.338  & & -0.956  & 0.202  & -0.466  & -2.120\Cross & -0.965  & & 0.397  & 0.651  & -0.677  & -1.395\Cross & -2.281\Cross \\
\cmidrule{3-7}  \cmidrule{9-13} \cmidrule{15-19}
LQR2& $\hat{\alpha}$&0.046 & 0.112 & 0.514 & 0.892 & 0.948 & & 0.050 & 0.108 & 0.494 & 0.900 & 0.950 & & 0.058 & 0.102 & 0.498 & 0.894 & 0.944 \\
 & $\hat{L}$ & 0.171 & 0.287 & 0.651 & 0.287 & 0.169 & & 0.389 & 0.649 & 1.415 & 0.566 & 0.327 & & 0.586 & 0.993 & 2.091 & 0.804 & 0.450 \\
 \bottomrule
\end{tabular}
\caption{Relative performance of alternative out-of-sample forecasts of WTI Crude Oil futures return quantiles. For each model, quantile ($\alpha$) and forecasts horizon ($h$), we report the unconditional coverage ($\hat{\alpha}$), the value of the tick-loss function ($\hat{L}$) and the Diebold-Mariano test statistic for equal predictive accuracy with the linear quantile regression model LQR2 serving as the benchmark. We use $*$ to denote significantly less accurate forecasts and \Cross  to denote significantly more accurate forecasts with respect to the benchmark at the 5\% significance level.\label{tab:outofsampleretOil}}
\end{center}
\end{sidewaystable}

\clearpage


\begin{sidewaystable}[!h]
\begin{center}
\renewcommand{\baselinestretch}{1.0}
\begin{tabular}{lcccccccccccccc}
\toprule
           & \multicolumn{4}{c}{HARQ1} & & \multicolumn{4}{c}{HARQ2} & & \multicolumn{4}{c}{HARQ3}\\
           \cmidrule{2-5}  \cmidrule{7-10}  \cmidrule{12-15}
\multicolumn{1}{r}{$\alpha$} & 0.50 & 0.75 & 0.90 & 0.95 & & 0.50 & 0.75 & 0.90 & 0.95 & & 0.50 & 0.75 & 0.90 & 0.95 \\
\midrule
\multicolumn{15}{l}{A. Parameter estimates} \\
const      & $\underset{(5.55)}{0.08}$ & $\underset{(4.23)}{0.07}$ & $\underset{(2.77)}{0.06}$ & $\underset{(1.31)}{0.05}$ & & $\underset{(-0.70)}{-0.01}$ & $\underset{(-0.46)}{-0.01}$ & $\underset{(-2.84)}{-0.07}$ & $\underset{(-2.68)}{-0.11}$ & & $\underset{(-1.32)}{-0.02}$ & $\underset{(-1.91)}{-0.04}$ & $\underset{(-3.36)}{-0.10}$ & $\underset{(-3.71)}{-0.15}$ \\
$RV_t^{1/2}$    & $\underset{(9.33)}{0.37}$ & $\underset{(9.31)}{0.50}$ & $\underset{(11.1)}{0.70}$ & $\underset{(7.69)}{0.78}$ & & & & & & & & &  \\
$RS_t^{{+}^{1/2}}$  & & & & & & $\underset{(-0.59)}{-0.01}$ & $\underset{(-0.93)}{-0.05}$ & $\underset{(-0.26)}{-0.02}$ & $\underset{(0.15)}{0.01}$ & & & & &  \\
$RS_t^{{-}^{1/2}}$  & & & & & & $\underset{(11.3)}{0.41}$ & $\underset{(10.1)}{0.58}$ & $\underset{(8.70)}{0.68}$ & $\underset{(6.21)}{0.76}$ & & & & &  \\
$RV_{t,t-5}^{1/2}$  & $\underset{(6.01)}{0.32}$ & $\underset{(6.67)}{0.40}$ & $\underset{(4.23)}{0.43}$ & $\underset{(3.04)}{0.49}$ & & $\underset{(7.13)}{0.27}$ & $\underset{(6.99)}{0.34}$ & $\underset{(5.75)}{0.43}$ & $\underset{(2.90)}{0.35}$ & & & & &  \\
$RV_{t,t-22}^{1/2}$ & $\underset{(4.37)}{0.16}$ & $\underset{(2.85)}{0.12}$ & $\underset{(1.41)}{0.11}$ & $\underset{(0.84)}{0.12}$ & & $\underset{(-1.18)}{-0.04}$ & $\underset{(-1.16)}{-0.04}$ & $\underset{(-2.85)}{-0.24}$ & $\underset{(-2.33)}{-0.23}$ & & & & &  \\
$IV_t^{1/2}$  &  & & & & & & & & & & $\underset{(10.8)}{0.30}$ & $\underset{(7.23)}{0.43}$ & $\underset{(7.04)}{0.53}$ & $\underset{(6.17)}{0.63}$ \\
$IV_{t,t-5}^{1/2}$  & & & & & & & & & & & $\underset{(5.44)}{0.22}$ & $\underset{(4.76)}{0.27}$ & $\underset{(4.29)}{0.31}$ & $\underset{(2.16)}{0.26}$ \\
$IV_{t,t-22}^{1/2}$   & & & & & & & & & & & $\underset{(-1.22)}{-0.04}$& $\underset{(-2.74)}{-0.13}$ & $\underset{(-2.61)}{-0.24}$ & $\underset{(-2.45)}{-0.27}$ \\
$JV_t^{1/2}$   & & & & & & & & & & & $\underset{(0.50)}{0.03}$ & $\underset{(1.40)}{0.14}$ & $\underset{(0.60)}{0.12}$ & $\underset{(1.82)}{0.46}$ \\
$VIX_t$   & & & & & & $\underset{(9.03)}{0.38}$ & $\underset{(6.94)}{0.38}$ & $\underset{(6.49)}{0.62}$ & $\underset{(6.93)}{0.75}$ & & $\underset{(10.6)}{0.41}$ & $\underset{(9.01)}{0.50}$ & $\underset{(6.89)}{0.71}$ & $\underset{(7.39)}{0.84}$ \\
\bottomrule
\end{tabular}
\caption{Estimated quantile regressions for the S\&P500 futures realized volatility $RV_{t+1}^{1/2}$. The table reports estimated coefficients with bootstrapped t-statistics in parentheses. The sample periods runs from January 3, 1997 till June 30, 2008.\label{tab:qrRVSP}}
\end{center}
\end{sidewaystable}

\clearpage


\begin{sidewaystable}[!h]
\begin{center}
\renewcommand{\baselinestretch}{1.0}
\begin{tabular}{lccccccccccccccc}
\toprule
           & \multicolumn{4}{c}{HARQ1} & & \multicolumn{4}{c}{HARQ2} & & \multicolumn{4}{c}{HARQ3} \\
           \cmidrule{2-5}  \cmidrule{7-10}  \cmidrule{12-15}
\multicolumn{1}{r}{$\alpha$} & 0.50 & 0.75 & 0.90 & 0.95 & & 0.50 & 0.75 & 0.90 & 0.95 & & 0.50 & 0.75 & 0.90 & 0.95 \\
\midrule
const          & $\underset{(3.32)}{0.15}$ & $\underset{(1.80)}{0.14}$ & $\underset{(2.33)}{0.23}$ & $\underset{(0.79)}{0.16}$ & & $\underset{(1.20)}{0.06}$ & $\underset{(-0.61)}{-0.05}$ & $\underset{(0.32)}{0.04}$ & $\underset{(-1.24)}{-0.24}$ & &
$\underset{(0.98)}{0.05}$ & $\underset{(-0.56)}{-0.04}$ & $\underset{(0.05)}{0.01}$ & $\underset{(-1.09)}{-0.22}$ \\
$D_t^{W}$      & $\underset{(8.95)}{0.23}$ & $\underset{(8.80)}{0.32}$ & $\underset{(6.50)}{0.31}$ & $\underset{(2.34)}{0.23}$ & & $\underset{(6.57)}{0.21}$ & $\underset{(7.01)}{0.30}$ & $\underset{(6.55)}{0.32}$ & $\underset{(3.45)}{0.28}$ & &
$\underset{(6.37)}{0.20}$ & $\underset{(7.13)}{0.30}$ & $\underset{(5.82)}{0.33}$ & $\underset{(2.68)}{0.26}$ \\
$RV_t^{1/2}$         & $\underset{(6.37)}{0.20}$ & $\underset{(3.43)}{0.23}$ & $\underset{(4.05)}{0.29}$ & $\underset{(1.37)}{0.22}$ & & & & & & & & & & \\
$RS_t^{{+}^{1/2}}$     & & & & & & $\underset{(1.59)}{0.05}$ & $\underset{(1.32)}{0.08}$ & $\underset{(0.54)}{0.04}$ & $\underset{(-0.15)}{-0.01}$ & & & & & \\
$RS_t^{{-}^{1/2}}$     & & & & & & $\underset{(4.69)}{0.16}$ & $\underset{(3.23)}{0.15}$ & $\underset{(1.56)}{0.14}$ & $\underset{(1.64)}{0.17}$ & & & & & \\
$RV_{t,t-5}^{1/2}$   & $\underset{(5.58)}{0.32}$ & $\underset{(6.29)}{0.43}$ & $\underset{(3.52)}{0.51}$ & $\underset{(3.02)}{0.81}$ & & $\underset{(4.63)}{0.31}$ & $\underset{(3.58)}{0.29}$ & $\underset{(3.72)}{0.52}$ & $\underset{(2.02)}{0.53}$ & & & & & \\
$RV_{t,t-22}^{1/2}$  & $\underset{(2.09)}{0.16}$ & $\underset{(2.06)}{0.15}$ & $\underset{(-0.36)}{-0.04}$ & $\underset{(0.07)}{0.01}$ & & $\underset{(5.41)}{0.31}$ & $\underset{(4.32)}{0.31}$ & $\underset{(1.93)}{0.22}$ & $\underset{(1.36)}{0.29}$ & & & & & \\
$IV_t^{1/2}$         & & & & & & & & & & & $\underset{(4.50)}{0.16}$ & $\underset{(3.95)}{0.19}$ & $\underset{(2.11)}{0.18}$ & $\underset{(1.44)}{0.19}$ \\
$IV_{t,t-5}^{1/2}$  & & & & & & & & & & & $\underset{(4.39)}{0.31}$ & $\underset{(3.38)}{0.27}$ & $\underset{(3.60)}{0.52}$ & $\underset{(1.92)}{0.50}$ \\
$IV_{t,t-22}^{1/2}$ & & & & & & & & & & & $\underset{(2.42)}{0.17}$ & $\underset{(2.03)}{0.15}$ & $\underset{(-0.74)}{-0.08}$ & $\underset{(0.19)}{0.03}$ \\
$JV_t^{1/2}$      & & & & & & & & & & & $\underset{(1.05)}{0.06}$ & $\underset{(0.47)}{0.02}$ & $\underset{(0.75)}{0.04}$ & $\underset{(-0.31)}{-0.03}$ \\
$ImV_t$      & & & & & & $\underset{(4.88)}{0.23}$ & $\underset{(5.50)}{0.41}$ & $\underset{(4.01)}{0.51}$ & $\underset{(4.26)}{0.72}$ & & $\underset{(4.25)}{0.21}$ & $\underset{(5.44)}{0.40}$ & $\underset{(4.16)}{0.52}$ & $\underset{(4.15)}{0.65}$ \\
\bottomrule
\end{tabular}
\caption{Estimated quantile regressions for the WTI Crude Oil futures realized volatility $RV_{t+1}^{1/2}$. The table reports estimated coefficients with bootstrapped t-statistics in parentheses. The sample periods runs from September 4, 2001 till August 30, 2008.\label{tab:qrRVOil}}
\end{center}
\end{sidewaystable}

\clearpage


\begin{table}[!h]
\begin{center}
\renewcommand{\baselinestretch}{1.0}
\begin{tabular}{lrrrrrrrrrr}
\toprule
  & & \multicolumn{4}{c}{in-sample} & & \multicolumn{4}{c}{out-of-sample} \\
\cmidrule{3-6}  \cmidrule{8-11}
 & $\alpha$ & 0.5 & 0.75 & 0.90 & 0.95 & & 0.5 & 0.75 & 0.90 & 0.95 \\
\midrule
\multicolumn{11}{l}{A. S\&P 500}  \\
ARFIMA & $\hat{\alpha}$ & 0.534 & 0.780 & 0.907 & 0.948 && 0.522 & 0.837 & 0.954 & 0.978 \\
       & $DQ$           & 25.08 & 29.13 & 24.53 & 33.08 && 46.73 & 51.52 & 31.01 & 15.67 \\
       & $p$-val        & 0.001 & 0.000 & 0.000 & 0.000 && 0.000 & 0.000 & 0.000 & 0.007 \\
\cmidrule{3-6}  \cmidrule{8-11}
LQR1 & $\hat{\alpha}$ & 0.500 & 0.750 & 0.900 & 0.949 && 0.554 & 0.770 & 0.894 & 0.944 \\
     & $DQ$           & 20.31 & 12.65 & 2.764 & 6.430 && 19.50 & 13.98 & 4.617 & 3.249 \\
     & $p$-val        & 0.002 & 0.045 & 0.820 & 0.405 && 0.002 & 0.028 & 0.613 & 0.859 \\
\cmidrule{3-6}  \cmidrule{8-11}
LQR2 & $\hat{\alpha}$ & 0.500 & 0.750 & 0.900 & 0.950 && 0.540 & 0.750 & 0.864 & 0.928 \\
     & $DQ$           & 18.93 & 25.25 & 5.369 & 4.877 && 12.99 & 6.462 & 10.09 & 6.633 \\
     & $p$-val        & 0.005 & 0.000 & 0.486 & 0.577 && 0.034 & 0.369 & 0.133 & 0.482 \\
\cmidrule{3-6}  \cmidrule{8-11}
LQR3 & $\hat{\alpha}$ & 0.500 & 0.750 & 0.900 & 0.950 && 0.538 & 0.750 & 0.862 & 0.928 \\
     & $DQ$           & 36.50 & 29.10 & 5.574 & 6.529 && 9.771 & 11.69 & 9.858 & 7.796 \\
     & $p$-val        & 0.000 & 0.000 & 0.475 & 0.368 && 0.134 & 0.068 & 0.178 & 0.334 \\
\midrule
\multicolumn{9}{l}{B. WTI Crude Oil} \\
ARFIMA & $\hat{\alpha}$ & 0.553 & 0.781 & 0.892 & 0.937 && 0.510 & 0.825 & 0.948 & 0.974 \\
       & $DQ$           & 47.61 & 26.63 & 22.36 & 13.97 && 13.33 & 21.03 & 19.52 & 13.71 \\
       & $p$-val        & 0.000 & 0.000 & 0.003 & 0.040 && 0.039 & 0.002 & 0.003 & 0.028 \\
\cmidrule{3-6}  \cmidrule{8-11}
LQR1   & $\hat{\alpha}$ & 0.501 & 0.750 & 0.901 & 0.949 && 0.524 & 0.734 & 0.884 & 0.948 \\
       & $DQ$           & 8.308 & 8.639 & 0.935 & 4.332 && 7.497 & 5.594 & 1.929 & 8.149 \\
       & $p$-val        & 0.207 & 0.189 & 0.989 & 0.649 && 0.274 & 0.486 & 0.923 & 0.278 \\
\cmidrule{3-6}  \cmidrule{8-11}
LQR2   & $\hat{\alpha}$ & 0.501 & 0.750 & 0.900 & 0.950 && 0.520 & 0.736 & 0.890 & 0.946 \\
       & $DQ$           & 11.79 & 9.121 & 1.729 & 1.954 && 7.263 & 7.053 & 3.484 & 3.976 \\
       & $p$-val        & 0.070 & 0.165 & 0.952 & 0.941 && 0.299 & 0.325 & 0.733 & 0.790 \\
\cmidrule{3-6}  \cmidrule{8-11}
LQR3   & $\hat{\alpha}$ & 0.501 & 0.750 & 0.900 & 0.949 && 0.524 & 0.730 & 0.900 & 0.950 \\
       & $DQ$           & 11.68 & 9.313 & 1.593 & 2.162 && 9.959 & 5.950 & 5.085 & 12.80 \\
       & $p$-val        & 0.071 & 0.175 & 0.950 & 0.916 && 0.148 & 0.435 & 0.559 & 0.033 \\
\bottomrule
\end{tabular}
\caption{Absolute performance of alternative conditional quantile models for daily S\&P500 and WTI Crude Oil futures realized volatility. The left-hand side panel reports results for in-sample performance and the right-hand side panel reports results for out-of-sample performance (one-step-ahead forecasts). For each model and quantile ($\alpha$) we report the unconditional coverage ($\hat{\alpha}$), the Berkowitz et al. (2011) test statistic for correct dynamic specification ($DQ$) and the corresponding Monte Carlo-based p-value ($p$-val).\label{tab:insampleRV}}
\end{center}
\end{table}

\clearpage


\begin{sidewaystable}[!h]
\begin{center}
\renewcommand{\tabcolsep}{1.8mm}
\renewcommand{\baselinestretch}{1.0}
\begin{tabular}{lrrrrrrrrrrrrrrr}
\toprule
  &&\multicolumn{4}{c}{h=1} & & \multicolumn{4}{c}{h=5} & & \multicolumn{4}{c}{h=10} \\
\cmidrule{3-6}\cmidrule{8-11}\cmidrule{13-16}
 & $\alpha$ & 0.5 & 0.75 & 0.90 & 0.95 & & 0.5 & 0.75 & 0.90 & 0.95 & & 0.5 & 0.75 & 0.90 & 0.95 \\
\midrule
\multicolumn{4}{l}{A. S\&P 500} \\
ARFIMA & $\hat{\alpha}$ & 0.522 & 0.837 & 0.954 & 0.978 && 0.550 & 0.745 & 0.849 & 0.902
                       && 0.556 & 0.723 & 0.823 & 0.865 \\
       & $\hat{L}$      & 0.051 & 0.047 & 0.029 & 0.018 && 0.066 & 0.063 & 0.043 & 0.029
                       && 0.073 & 0.073 & 0.053 & 0.038 \\
       & $DM$           & -15.66\Cross & -10.84\Cross & -6.093\Cross & -4.286\Cross
                       && -1.852\Cross & -0.391  & 0.003  & -0.307
                       && -0.257  & 0.648  & 0.911 & 1.025 \\
\cmidrule{3-6}\cmidrule{8-11}\cmidrule{13-16}
LQR1 & $\hat{\alpha}$ & 0.554 & 0.770 & 0.894 & 0.944 && 0.582 & 0.734 & 0.882 & 0.926
                     && 0.584 & 0.752 & 0.858 & 0.908 \\
     & $\hat{L}$      & 0.079 & 0.072 & 0.044 & 0.029 && 0.076 & 0.068 & 0.045 & 0.030
                     && 0.080 & 0.075 & 0.049 & 0.032 \\
     & $DM$           & 4.307* & 2.437* & 1.559 & 2.224* && 4.606* & 2.812* & 1.580 & 1.227
                     && 2.645* & 2.138* & 1.079 & 0.999 \\
\cmidrule{3-6}\cmidrule{8-11}\cmidrule{13-16}
LQR2 & $\hat{\alpha}$ & 0.540 & 0.750 & 0.864 & 0.928 && 0.536 & 0.732 & 0.870 & 0.910
                     && 0.558 & 0.740 & 0.848 & 0.898 \\
     & $\hat{L}$      & 0.074 & 0.068 & 0.042 & 0.027 && 0.069 & 0.064 & 0.042 & 0.029
                     && 0.074 & 0.070 & 0.047 & 0.031 \\
     & $DM$           & -1.707 & -1.228 & -0.415 & 0.477 && -1.796 & -0.954 & -0.578 & -1.189
                     && 0.684 & -0.651 & -0.649 & -0.583 \\
\cmidrule{3-6}\cmidrule{8-11}\cmidrule{13-16}
LQr3 & $\hat{\alpha}$ & 0.538 & 0.750 & 0.862 & 0.928 && 0.536 & 0.734 & 0.870 & 0.920
                     && 0.568 & 0.730 & 0.842 & 0.900 \\
     & $\hat{L}$      & 0.075 & 0.069 & 0.042 & 0.026 && 0.070 & 0.064 & 0.043 & 0.029
                     && 0.074 & 0.070 & 0.047 & 0.031 \\
\midrule
\multicolumn{4}{l}{B: WTI Crude Oil} \\
ARFIMA & $\hat{\alpha}$ & 0.510 & 0.825 & 0.948 & 0.974 && 0.488 & 0.783 & 0.904 & 0.944
                       && 0.528 & 0.747 & 0.873 & 0.902 \\
       & $\hat{L}$      & 0.096 & 0.085 & 0.053 & 0.033 && 0.078 & 0.069 & 0.042 & 0.026
                       && 0.081 & 0.074 & 0.045 & 0.027 \\
       & $DM$           & -9.949\Cross & -7.823\Cross & -5.159\Cross & -3.908\Cross
                       && -1.518\Cross & -0.148 & 0.719  & 0.486
                       && 0.349 & 1.089 & 1.036 & 0.649 \\
\cmidrule{3-6}\cmidrule{8-11}\cmidrule{13-16}
LQR1   & $\hat{\alpha}$ & 0.524 & 0.734 & 0.884 & 0.948 && 0.522 & 0.754 & 0.886 & 0.926
                       && 0.554 & 0.752 & 0.876 & 0.924 \\
       & $\hat{L}$      & 0.127 & 0.116 & 0.070 & 0.043 && 0.088 & 0.079 & 0.047 & 0.028
                       && 0.087 & 0.077 & 0.047 & 0.028 \\
       & $DM$           & 1.070 & 2.605* & 2.426* & 1.730* && 1.119 & 2.285* & 2.094* & 1.473
                       && 2.118* & 1.907* & 1.622 & 1.057 \\
\cmidrule{3-6}\cmidrule{8-11}\cmidrule{13-16}
LQR2   & $\hat{\alpha}$ & 0.520 & 0.736 & 0.890 & 0.946 && 0.520 & 0.720 & 0.902 & 0.952
                       && 0.552 & 0.742 & 0.902 & 0.940 \\
       & $\hat{L}$      & 0.126 & 0.113 & 0.066 & 0.041 && 0.086 & 0.070 & 0.040 & 0.024
                       && 0.080 & 0.068 & 0.039 & 0.026 \\
       & $DM$           & 0.933 & 2.295 & 0.525 & 0.478 && 1.552 & 0.528 & 0.513 & 0.142
                       && 0.727 & -0.386 & -1.274 & 1.793 \\
\cmidrule{3-6}\cmidrule{8-11}\cmidrule{13-16}
LQR3   &$\hat{\alpha}$  & 0.524 & 0.730 & 0.900 & 0.950 && 0.516 & 0.728 & 0.898 & 0.950
                       && 0.548 & 0.742 & 0.894 & 0.944 \\
       & $\hat{L}$      & 0.126 & 0.111 & 0.066 & 0.040 && 0.085 & 0.070 & 0.040 & 0.024
                       && 0.079 & 0.068 & 0.039 & 0.025 \\
\bottomrule
\end{tabular}
\caption{Relative performance of alternative out-of-sample forecasts of S\&P500 and WTI Crude Oil futures realized volatility quantiles. For each model, quantile ($\alpha$) and forecasts horizon ($h$), we report the unconditional coverage ($\hat{\alpha}$), the value of the tick-loss function ($\hat{L}$) and the Diebold-Mariano test statistic for equal predictive accuracy with the linear quantile regression model LQR2 serving as the benchmark. We use $*$ to denote significantly less accurate forecasts and \Cross to denote significantly more accurate forecasts with respect to the benchmark at the 5\% significance level.\label{tab:outofsampleRV}}
\end{center}
\end{sidewaystable}

\clearpage

\section*{Figures}

\clearpage

\clearpage


\clearpage

\end{document}